\documentclass{aa}
\usepackage[varg]{txfonts}
\usepackage{graphicx}
\usepackage{natbib}
\bibpunct{(}{)}{;}{a}{}{,}

\usepackage[english]{babel}

\usepackage{hyperref}
\hypersetup{colorlinks=true,
 linkcolor=black,
 citecolor=black,
 filecolor=black,
 urlcolor=black
}

\begin{document}

\title{TRADES: A new software to derive orbital parameters from observed transit times and radial velocities}
\titlerunning{TRADES: a new software to derive orbital parameters}

\subtitle{Revisiting Kepler-11 and Kepler-9}

\author{L. Borsato\inst{1,2}
  \and F. Marzari\inst{1}
  \and V. Nascimbeni\inst{1,2}
  \and G. Piotto\inst{1,2}
  \and V. Granata\inst{1,2}
  \and L. R. Bedin\inst{2}
  \and L. Malavolta\inst{1,2}
}

\institute{Department of Physics and Astronomy, Universit\`a degli Studi di Padova, Via Marzolo, 8 I-35131 Padova\\
  \email{luca.borsato.2@studenti.unipd.it}
  \and INAF – Osservatorio Astronomico di Padova, vicolo dell’Osservatorio 5, 35122 Padova, Italy
}

\date{Received 28 April, 2014 / Accepted 7 August, 2014}

\abstract{}
	{With the purpose of determining the orbital parameters of exoplanetary systems from observational data, we have developed a software, named \texttt{TRADES} (TRAnsits and Dynamics of Exoplanetary Systems), to simultaneously fit observed radial velocities and transit times data.}
	{We implemented a dynamical simulator for $N$-body systems, which also fits the available data during the orbital integration and determines the best combination of the orbital parameters using grid search, $\chi^2$ minimization, genetic algorithms, particle swarm optimization, and bootstrap analysis.}
	{To validate \texttt{TRADES}, we tested the code on a synthetic three-body system and on two real systems discovered by the \textit{Kepler} mission: Kepler-9 and Kepler-11.
	These systems are good benchmarks to test multiple exoplanet systems showing transit time variations (TTVs) due to the gravitational interaction among planets.
	We have found that orbital parameters of Kepler-11 planets agree well with the values proposed in the discovery paper and with a a recent work from the same authors.
	We analyzed the first three quarters of Kepler-9 system and found parameters in partial agreement with discovery paper.
	Analyzing transit times ($T_0$s), covering 12 quarters of \textit{Kepler} data, that we have found a new best-fit solution.
	This solution outputs masses that are about $55\%$ of the values proposed in the discovery paper; this leads to a reduced semi-amplitude of the radial velocities of about $12.80$ m$s^{-1}$.
	}
	{}


\maketitle

\section{Introduction}
\label{intro}

Now, more than $1779$ planets\footnote{\href{http://exoplanet.eu/catalog/}{http://exoplanet.eu/catalog}, 2014 March $21^{\mathrm{st}}$.} have been discovered and confirmed in about 1102 planetary systems.
Around 460 planetary systems are known to be multiple planet systems.
Hundreds of \textit{Kepler} planetary candidates with multiple transit-like signals are still waiting confirmation \citep[see][]{Latham2011ApJ,Lissauer2011ApJ}.
The usual way to characterize multiple planet systems is by combining information from both transits and radial velocities (RVs).
An effect due to the presence of multiple planets is the transit time variation (TTV): the gravitational interaction between two planets causes a deviation from the Keplerian orbit and, as a consequence, the transit times ($T_0$s) of a planet may be not strictly periodic \citep[see][]{Agol2005,HM2005,Mi-Es2002}.
This effect can be also exploited to infer the presence of an unknown planet, even if it does not transit the host star \citep{Agol2005,HM2005}.
For example, the \textit{Kepler} Transit Timing Observations series \citep[TTO,][and references therein]{Ford2011ApJS} and TASTE project \citep[][and references therein]{Nascimbeni2011a} demonstrate the use of this technique.
\par
The problem of determining the masses and the orbital parameters of the planets in a multiple system is a difficult inverse problem.
In some works, the authors adopted an analytic approach to the problem \citep[e.g.,\ ][]{NesMor2008ApJ,Nesv2009}, developing a method from the perturbation theory \citep{Hori1966,Deprit1969}, where the $T_0$s are computed as a Fourier series.
A drawback of this method is that it does not take into account for the so-called mean motion resonance (MMR) or cases that are just outside the MMR.
We have a MMR when the ratio of the periods of two planets is a multiple of a small integer number, such as 2:1.
Two planets reciprocally in MMR, or just outside it, show a strong TTV signal that is easily detectable even with ground-base facilities.\\
The method described in this paper is based on a direct numerical \textit{N}-body approach \citep{StAg2005,AgSt2007}, which is conceptually simpler, but computationally intensive.
Very recently, \citet{Deck2014} have developed TTVFast, a symplectic integrator that computes transit times and radial velocities of an exoplanetary system.\\
An example of the application based on the TTV technique can be found in \citet{Nesv2013ApJ}, where the authors have predicted the presence of the planet KOI-142c in the system, which has been recently confirmed by \citet{Barros2013arXiv}.
\par

In Sect.~\ref{trades} we introduce the \texttt{TRADES} program, the basic formulas, and methods to calculate radial velocities and transit times; in Sects.~\ref{simulated}, \ref{kepler-11}, and \ref{kepler-9}, we run \texttt{TRADES} on a synthetic \mbox{$3$-body} system, Kepler-11, and Kepler-9, respectively. We summarize and discuss the results in the Sect.~\ref{summary}.\par

\section{TRADES}
\label{trades}

We have developed a computer program (in Fortran 90 and openMP) for determining the possible physical and dynamical configurations of extra-solar planetary systems from observational data, known as \texttt{TRADES}, which stands for TRAnsits and Dynamics of Exoplanetary Systems.
The program \texttt{TRADES} models the dynamics of multiple planet systems and reproduces the observed transit times ($T_{0}$, or mid-transit times) and radial velocities (RVs).
These $T_{0}$s and RVs are computed during the integration of the planetary orbits.
We have developed \texttt{TRADES} from zero because we want to avoid black-box programs,it would be easier to parallelize it with openMP, and include additional algorithms.
\par

To solve the inverse problem, \texttt{TRADES} can be run in four different modes:
1) `grid' search,
2) Levenberg-Marquardt\footnote{\texttt{lmdif} converted to \texttt{Fortran~90} by \href{http://jblevins.org/mirror/amiller/}{Alan Miller (http://jblevins.org/mirror/amiller/)} from \texttt{MINPACK}.} (\texttt{LM}) algorithm,
3) genetic algorithm \citep[\texttt{GA}, we used the implementation named \texttt{PIKAIA}\footnote{\href{http://www.hao.ucar.edu/modeling/pikaia/pikaia.php}{PIKAIA (http://www.hao.ucar.edu/modeling/pikaia/pikaia.php)} converted to \texttt{Fortran 90} by \href{http://jblevins.org/mirror/amiller/}{Alan Miller}.},][]{Charbonneau1995PIK},
4) and particle swarm optimization \citep[\texttt{PSO}\footnote{based on the public \texttt{Fortran~90} code at \href{http://www.nda.ac.jp/cc/users/tada/}{http://www.nda.ac.jp/cc/users/tada/}},][]{PSO_TADAT}.
In each mode, \texttt{TRADES} compares observed transit times ($T_{0,\mathrm{obs}}$s) and radial velocities (RV$_{\mathrm{obs}}$s) with the simulated ones ($T_{0,\mathrm{sim}}$s and RV$_{\mathrm{sim}}$s).\par

\begin{enumerate}

\item In the grid search method, \texttt{TRADES} samples the orbital elements of a perturbing body in a four-dimensional grid:
the mass, $M$, the period, $P$ (or the semi-major axis, $a$), the eccentricity, $e$, and the argument of the pericenter, $\omega$.
The grid parameters can be evenly sampled on a fixed grid by setting the number of steps, or the step size, or by a number of points chosen randomly within the parameter bounds.
For any given set of values, the orbits are integrated, and the residuals between the observed and computed $T_0$s and RVs are computed.
For each combination of the parameters, the \texttt{LM} algorithm can be called and the best case is the one with the lowest residuals (lowest $\chi^{2}$).
We have selected these four parameters for the grid search because they represent the minimal set of parameters required to model a coplanar system.
In the future, we intend to add the possibility of making the grid search over all the set of parameters for each body.
\par

\item After an initial guess on the orbital parameters of the perturber, which could be provided by the previously described grid approach, the \texttt{LM} algorithm exploits the Levenberg-Marquardt minimization method to find the solution with the lowest residuals.
The \texttt{LM} algorithm requires the analytic derivative of the model with respect to the parameters to be fitted.
Since the $T_0$s are determined by an iterative method and the radial velocities are computed using the numerical integrator, we cannot express these as analytic functions of fitting parameters.
We have adopted the method described in \citet{MINPACK-1} to compute the Jacobian matrix, which is determined by a forward-difference approximation.
The \texttt{epsfcn} parameter, which is the parameter that determines the first Jacobian matrix, is automatically selected in a logarithmic range from the machine precision up to $10^{-6}$;
the best value is the one that returns the lower $\chi^2$.
This method has the advantage to be scale invariant, but it assumes that each parameter is varied by the same \texttt{epsfcn} value (e.g.,\ a variation of $10\%$ of the period has a different effect than a variation of the same percentage of the argument of pericenter).\par

\item The \texttt{GA} mode searches for the best orbit by performing a genetic optimization \citep[e.g.][]{GA_01,GA_02}, where the fitness parameter is set to the inverse of the $\chi^{2}$.
This algorithm is inspired by natural selection which is the biological process of evolution.
Each generation is a new population of `offspring' orbital configurations, that are the result of `parent' pairs of orbital configurations that are ranked following the fitness parameter.
A drawback of the \texttt{GA} is the slowness of the algorithm, when compared to other optimizers.
However, the \texttt{GA} should converge to a global solution (if it exists) after the appropriate number of iterations.

\item The \texttt{PSO} is another optimization algorithm that searches for the global solution of the problem;
this approach is inspired by the social behavior of bird flock and fish school \citep[e.g., ][]{PSO_01,PSO_02}.
The fitness parameter used is the same as the \texttt{GA}, the inverse of the $\chi^{2}$.
For each `particle', the next step (or iteration) in the space of the fitted parameters is mainly given by the combination of three terms:
random walk, best `particle' position (combination of parameters), and best `global' position (best orbital configuration of the all particles and all iterations).

\end{enumerate}

The grid search is a good approach in case that we want to explore a limited subset of the parameter space or if we want to analyze the behavior of the system by varying some parameters, for example to test the effects of a growing mass for the perturbing planet.
\texttt{GA} and \texttt{PSO} are good methods to be used in case of a wider space of parameters.
The orbital solution determined with the \texttt{GA} or the \texttt{PSO} method is eventually refined with the \texttt{LM} mode.\par

For each mode, \texttt{TRADES} can perform a bootstrap analysis to calculate the interval of confidence of the best-fit parameter set.
We generate a set of $T_0$s and RVs from the fitted parameters, and we add a Gaussian noise having the calculated value (of $T_0$s and RVs) as the mean and the corresponding measurement error as variance.
We fit each new set of observables with the \texttt{LM}.
We iterate the whole process thousands of times to analyze the distribution for each fitted parameter.\par

\subsection{Orientation of the reference frame}
\label{refframe}

For the transit time determination, the propagation of the trajectories of all planets in the system is performed in a reference frame with the $Z$-axis pointing to the observer, while the $X$-$Y$ plane is the sky plane.
At a given reference epoch, the Keplerian orbital elements of each planet are:
period $P$ (or semi-major axis $a$),
inclination $i$,
eccentricity $e$,
argument of the pericenter $\omega$,
longitude of the ascending node $\Omega$,
and time of the passage at the pericenter $\tau$ (or the mean anomaly M).
Given the orbital elements we first compute the initial radius $\vec r$ and velocity $\dot{\vec r}$ vectors in the orbital plane \citep[e.g.,\ see ][]{MuDe2000book}:
\begin{equation}
  \label{eq:rorbit}
  \begin{array}{c}
    \\
    \vec r = \\
    \\
  \end{array}
  \left(\begin{array}{c}
    x\\
    y\\
    z\\
  \end{array}\right)
  =
  \left(\begin{array}{c}
    a \left( \cos E - e \right)\\
    a \sqrt{1-e^2} \sin E\\
    0
  \end{array}\right)\; ,
\end{equation}
\begin{equation}
  \label{eq:vorbit}
  \begin{array}{c}
    \\
    \dot{\vec r }= \\
    \\
  \end{array}
  \left(\begin{array}{c}
    \dot{x}\\
    \dot{y}\\
    \dot{z}\\
  \end{array}\right)
  =
  \left(\begin{array}{c}
    \frac{n}{1-e \cos E} \left( -a \sin E \right)\\
    \frac{n}{1-e \cos E} \left( a \sqrt{1-e^2} \cos E \right)\\
    0
  \end{array}\right)\; ,
\end{equation}
where $n=2 \pi / P$ is the mean motion, $E$ is the eccentric anomaly obtained from the solution of the Kepler's equation, \mbox{$\mathrm{M} = E - e \sin E$}, with the Newton-Raphson method \citep[e.g.,\ see][]{Danby1988book,MuDe2000book,MuCo2011exop}.\\
Then, we rotate the state vector by applying three consecutive rotation matrices, $\textrm{R}_{l}(\phi)$ \citep[e.g.,\ see][]{Danby1988book,MuDe2000book,MuCo2011exop} where $\phi$ is the rotation angle and $l$ is the rotation axis (where $l$ is $\{1,2,3\}$ for $\{x^{\prime},y^{\prime},z^{\prime}\}$).
To rotate the initial state vector from the orbital plane to the observer reference frame, we have to use the transpose of the rotation matrix, $\textrm{R}^{\mathrm{T}}_{l}(\phi)$ with angles $\omega$, $i$, and $\Omega$.
After this rotation, the $X$-$Y$ plane is the sky plane with the $Z$-axis pointing to the observer, and we determine the initial state vector of each \textit{k-th} planet:
\begin{equation}
  \label{eq:rotation}
  \left(\begin{array}{c}
    X\\
    Y\\
    Z\\
  \end{array}\right)
  = \mathrm{R}^{\mathrm{T}}_{3}(\Omega) \, \mathrm{R}^{\mathrm{T}}_{1}(i) \, \mathrm{R}^{\mathrm{T}}_{3}(\omega)
  \left(\begin{array}{c}
    x\\
    y\\
    z\\
  \end{array}\right)\; .
\end{equation}
The same rotations have to be applied to the initial velocity vector.
The inclinations are measured from the sky-plane.
Indeed, a planet with inclination of $0\degr$ has an orbit that lies on the sky-plane ($X$-$Y$ plane), that is, it is seen face-on.
The orbit of a planet with $i=90\degr$ is seen edge-on (it transits exactly through the center of the star), and it lies on the $X$-$Z$ plane.
From the initial state vector, \texttt{TRADES} integrates the astrocentric equation of motion \citep[e.g.][]{MuDe2000book, Fab2011exop} of planet $k$:
\begin{equation}
  \label{eq:force}
  \ddot{\vec r}_{k} =
  -G\left(M_{1}+M_{k}\right) \frac{\vec r _{k}}{r^{3}_{k}}
  +G \sum^{N}_{j=2;j\ne k}{M_{j}\left(\frac{\vec r_{j} - \vec r_{k}}{|\vec r_{j} - \vec r_{k}|^{3}}-\frac{\vec r_{j}}{r^3_{j}}\right)}\; ,
\end{equation}
where $M_{1}$ is the mass of the star and $N$ the number of bodies; the first term is the \textit{direct} gravitational force, and the second term is the \textit{indirect} force due to mutual interaction of the planets.
The orbits are computed with the Runge-Kutta-Cash-Karp integrator \citep[RKCK,][]{RKCK1990,NR-Press1996}.
It is not a symplectic integrator\footnote{A symplectic integrator has been designed to numerically solve the Hamilton's equation by preserving the Poincar\'e invariants.}, and it is not well suited for long--term time integrations.
Instead, it uses small and variable steps (it self adjusts the step-size to maintain the numerical precision during the computation of the orbits), is fast, and preserves the total energy and the total angular momentum during the time scales of our simulations.
\par

\subsection{Transit determination}
\label{Tdet}

We chose the change of sign of the $X$ or $Y$ coordinates between two consecutive steps of each planet trajectory as first condition of an eclipse.
When this condition is met, following \citet[chap.~2.5]{Fab2011exop}, we have to seek roots of the sky-projected separation $\vec r_{s,k} \equiv (X_{k},Y_{k})$ with the Newton-Raphson method by solving
\begin{equation}
  \label{eq:gxy}
  g(X_{k},\dot{X_{k}},Y_{k},\dot{Y}_{k}) =
  \vec r_{s,k} \cdot \dot{\vec r}_{s,k} =
  X_{k}\dot{X}_{k}+Y_{k}\dot{Y}_{k} = 0
\end{equation}
then moving and iterating by the quantity,
\begin{equation}
  \label{eq:dtmid}
  \delta t = -g \left(\frac{\partial g}{\partial t}\right)^{-1}\; .
\end{equation}
In this way, we can determine with high precision the mid-transit time and the corresponding state vector ($\vec r_{\mathrm{mid}}$, $\dot{\vec r}_{\mathrm{mid}}$) with an accuracy equal to the selected $\delta t$.
We decided to set this accuracy in \texttt{TRADES} at the machine precision, which can be fine-tuned in the source of the code and defines the type of the chosen variables.
\par

Then, we determine if we have four contact times, or just two (in the case of a grazing eclipse), or no transit, comparing the module of the sky-projected separation at the transit time, $| \vec r_{s,{\mathrm{mid}}}|$, with the radius of the star, $R_{\star}$, and of the planets, $R_{k}$, as in \citet{Fab2011exop}.
If the transit (or the occultation) does exist, we move about $\mp R_{\star}/|\dot{\vec r}_{s,{\mathrm{mid}}}|$ from the $t_{\mathrm{mid}}$ backward ($-$) for first and second contact, and forward ($+$) for third and fourth contact.
Then, we solve
\begin{equation}
  \label{eq:hxy}
  h(X_{k},Y_{k}) =
  \dot{X}_{k}^{2}+X_{k}\ddot{X}_{k}+\dot{Y}_{k}^{2}+Y_{k}\ddot{Y}_{k} = 0
\end{equation}
and move of
\begin{equation}
  \label{eq:dtcont}
  \delta t = -h \left(\frac{\partial h}{\partial t}\right)^{-1}
\end{equation}
until $\delta t$ is less than the accuracy found (the same adopted in finding the transit time).
\par

We based our approach on \citet{Fab2011exop}, but we used a bisection-Newton-Raphson hybrid method, which is guaranteed to be bound near the solution,
and we assume that the orbital elements of the bodies are almost constant around the center of the transit.
Because of the latter assumption, we use $F(t_i,t_{i-1})$ and $G(t_i,t_{i-1})$ functions \citep[called $f(t,t_{0})$ and $g(t,t_{0})$ in ][]{Danby1988book,MuDe2000book}
to compute the planetary state vectors instead of the integrator while seeking the transit times:
\begin{equation}
  \label{eq:rvfg}
  \left\{ \begin{array}{l}
    \vec r_i(t) = F(t_i,t_{i-1})\vec r_{i-1} + G(t_i,t_{i-1})\dot{\vec r}_{i-1}\\
    \dot{\vec r}_i(t) = \dot{F}(t_i,t_{i-1})\vec r_{i-1} + \dot{G}(t_i,t_{i-1})\dot{\vec r}_{i-1}\\
  \end{array} \right.
\end{equation}
where
\begin{equation}
  \label{eq:riri-1}
  \left\{ \begin{array}{l}
    \vec r_{i-1} = \vec r(t_{i-1})\\
    \dot{\vec r}_{i-1} = \dot{\vec r}(t_{i-1})\\
  \end{array} \right.
\end{equation}
with
\begin{equation}
  \label{eq:f&g}
  \left\{ \begin{array}{l}
    F(t_i,t_{i-1})=\frac{a_{i}}{r_{i-1}} \left[ \cos (E_i-E_{i-1})-1 \right] +1\\
    G(t_i,t_{i-1})=(t_i-t_{i-1})+\frac{1}{n_i} \left[ \sin (E_i-E_{i-1}) - (E_i-E_{i-1}) \right]
  \end{array} \right.
\end{equation}
and
\begin{equation}
  \label{eq:fdot&gdot}
  \left\{ \begin{array}{l}
    \dot F(t_i,t_{i-1})=-\frac{a_i^{2}}{r_ir_{i-1}} n_i \sin (E_i-E_{i-1}) \\
    \dot G(t_i,t_{i-1})=\frac{a_i}{r_i} \left[ \cos (E_i-E_{i-1})-1 \right] +1\\
  \end{array} \right. \; .
\end{equation}
where $t_i = t_{i-1} + \delta t_{i-1}$ at the $i$-\textit{th} iterations; $E_i$ and $E_{i-1}$ are the eccentric anomalies at the $i$-\textit{th} and $i$-\textit{th}$-1$ iterations.
This allows the code to run faster than using the integrator, which has a lower number of function calls, to seek the transit and contact times.\par

The light coming from the star is delayed due to the motion of the star around the barycenter of the system \citep{Irwin1952}, and so \texttt{TRADES} corrects for the light-time travel effect \citep[LTE $= -Z_{\star}^{\mathrm{barycentric}}/c$, see][]{Fab2011exop}, each contact, and center transit time.\par

\subsection{RV calculation and other constraints}
\label{RVcalc}

For each observed RV (when available), \texttt{TRADES} integrates the orbits of the planets to the instant of the RV point and calculates the RV as the opposite of the $z$-component of the barycentric velocity of the star ($\mathrm{rv}_\mathrm{sim} = -\dot{Z}_{\star}^{\mathrm{barycentric}}$ in the right unit of measurement).
The observed RV is defined as $\mathrm{RV}_\mathrm{obs} = \gamma + \mathrm{rv}_\mathrm{obs}$, where $\gamma$ is the motion of the barycenter of the system and $\mathrm{rv}_\mathrm{obs}$ is the reflex motion of the star induced by the planets.
The program \texttt{TRADES} calculates $\gamma_\mathrm{sim}$ as the weighted mean of the difference $\Delta \mathrm{rv}_{j} = \mathrm{RV}_{j,\mathrm{obs}} - \mathrm{rv}_{j,\mathrm{sim}}$ with $j$ from one to the number of RVs.
The final simulated RV is $\mathrm{RV}_{j,\mathrm{sim}} = \gamma_\mathrm{sim} + \mathrm{rv}_{j,\mathrm{sim}}$.
We are planning to implement the $\gamma_\mathrm{sim}$ fitting rather than the described weighted mean method.
\par

Furthermore, we added some constraints on the orbit during the integration, setting a minimum and a maximum semi-major-axis for the system, $a_{\mathrm{min}}$ and $a_{\mathrm{max}}$, respectively.
The lower limit has been set equal to the star radius, while the maximum limit has been set equal to five times the largest semi-major axis of the system calculated from the periods of the planets.
In the \texttt{GA}, we have used the largest period boundary.
We use the definition of the Hill's sphere to obtain minimum distance allowed between two planets \citep{MuDe2000book}.
In this case, when these constraints are not be respected, the integration is stopped, and the $\chi^{2}$ returned is set to the maximum value allowed by the compiler, so the combination of the parameters is rejected.\par

\section{Validation with simulated system}
\label{simulated}

To validate \texttt{TRADES}, we simulated a synthetic system with two planets having known orbital parameters.
We chose a star with a mass and radius equal to the Sun, a first planet named b with Jupiter mass and radius ($M_{\mathrm{Jup}}$ and $R_{\mathrm{Jup}}$), and a second planet named c with mass and radius of Saturn ($M_{\mathrm{Sat}}$ and $R_{\mathrm{Sat}}$).
We assumed a co-planar system with inclination of $90\degr$ (perfectly edge-on).
The input orbital elements of the system are summarized in Table~\ref{Tabsynth}.

\begin{table}[h]
  \caption{
    Parameters of the simulated system in section \ref{simulated}.
  }
  \label{Tabsynth}
  \centering
  \small
  \begin{tabular}{c c c c}
    \hline\hline\noalign{\smallskip}
    Parameter & Star & Planet b & Planet c \\
    \hline\noalign{\smallskip}
    $M$ & $1.~M_{\sun}$ & $1.~M_{\mathrm{Jup}}$ & $1.~M_{\mathrm{Sat}}$\\
    $R$ & $1.~R_{\sun}$ & $1.~R_{\mathrm{Jup}}$ & $1.~R_{\mathrm{Sat}}$\\
    $a$ &              & $0.1$~AU            & $0.2$~AU         \\
    $e$ &              & $0.1$               & $0.3$             \\
    $\omega$ &         & $90\degr$           & $90\degr$         \\
    $\mathrm{M}$ &     & $0\degr$            & $0\degr$          \\
    $i$ &              & $90\degr$           & $90\degr$          \\
    $\Omega$ &         & $0\degr$            & $0\degr$          \\
    \noalign{\smallskip}\hline
  \end{tabular}
\end{table}

We simulated the system with \texttt{TRADES} for 500 days.
We computed all $T_0$s for each body and we call these times the `true' transit times ($T_{0,\mathrm{true}}$s) of the system.
Then, we created sets of synthetic transit times ($T_{0,\mathrm{synth}}$s), such as
\begin{equation}
  \label{eqT0synth}
  T_{0,\mathrm{synth}} = T_{0,\mathrm{true}} + N(0,1) \times \frac{P}{3} \times s \; ,
\end{equation}
where $s$ is a scaling factor varying from $0.01$ to $1.5$ (on twenty logarithmic steps) needed to simulate good to very bad measurement cases;
$N(0,1)$ is Gaussian noise with $0$ as mean and $1$ as variance.
The $P/3$ factor is needed to scale the Gaussian noise in the right unit of time, and avoids confusion between transits and occultations at the same time.
Furthermore, for each set of $T_{0,\mathrm{synth}}$s, we selected a random number of transits (at least $N/3$ with $N$ being the total number of transits of each planet) to simulate observed transits.\par

\begin{figure}[!]
  \resizebox{\hsize}{!}{\includegraphics{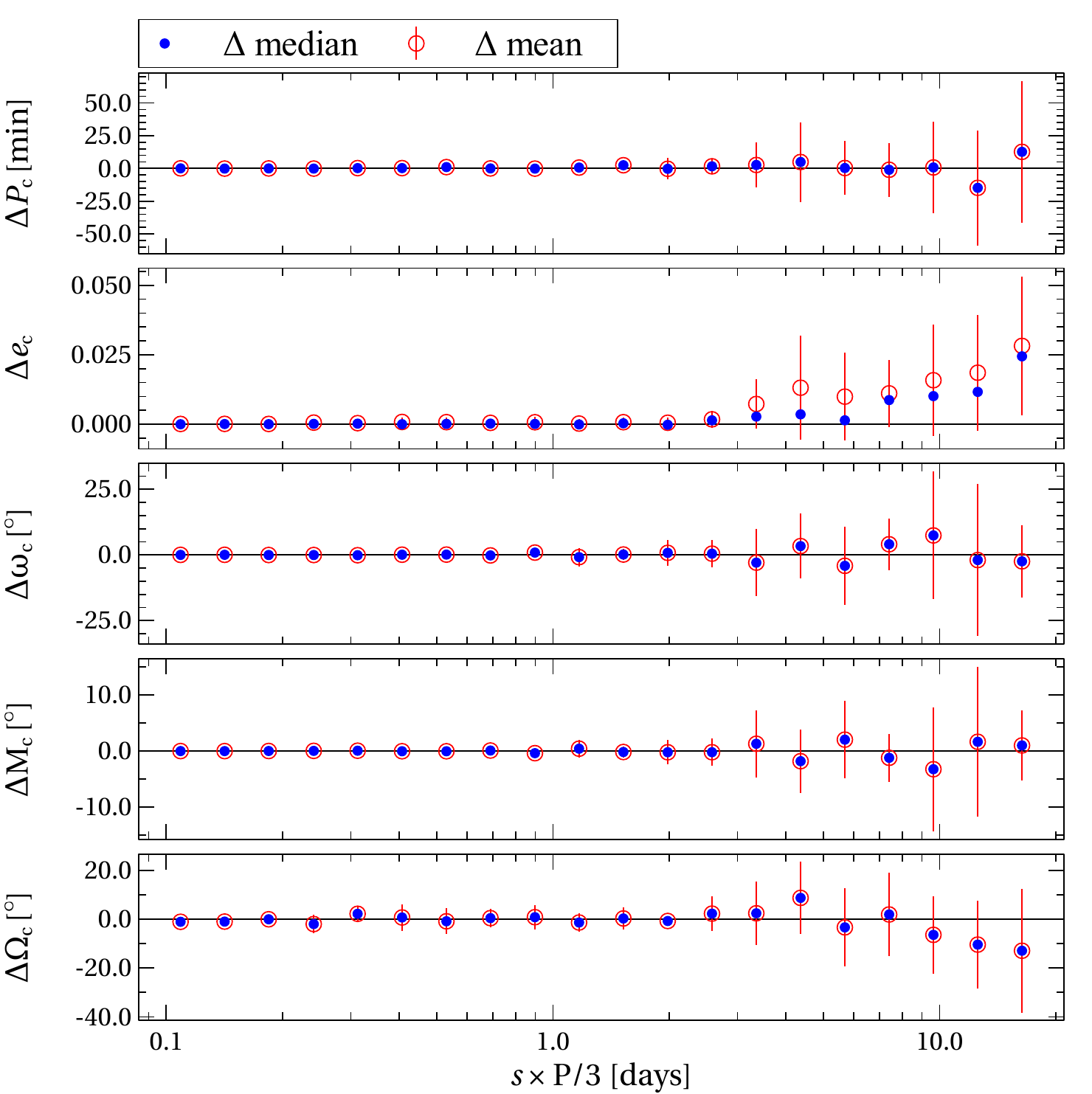}}
  \caption{
    Mean (red-open circles with $1\sigma$ error bars) and median (blue-filled circles) variation ($\Delta$) of fitted parameters of planet c for each value of the measurement errors on $T_0$, which are here parametrized as $s \times P/3$ (see text for details).
  }
  \label{FigSynth}
\end{figure}

We fixed the orbital parameters of planet b, and we fitted $M$, $P$, $e$, $\omega$, $\mathrm{M}$, and $\Omega$ of planet c.
We ran \texttt{TRADES} in \texttt{LM} mode for each scaling factor, and we calculated the difference of the parameters ($\Delta$) as the determined parameters minus the input parameters.
We repeated the simulation ten times (we calculated new Gaussian noise and the number of observed times every time).
In Fig.~\ref{FigSynth}, we plotted then mean and median of ten simulations for each $s$ scaling factor value.
The parameters of the system derived by \texttt{TRADES} depart from the `true' values only for extremely large measurement errors.
\par

\begin{figure}[!]
  \resizebox{\hsize}{!}{\includegraphics{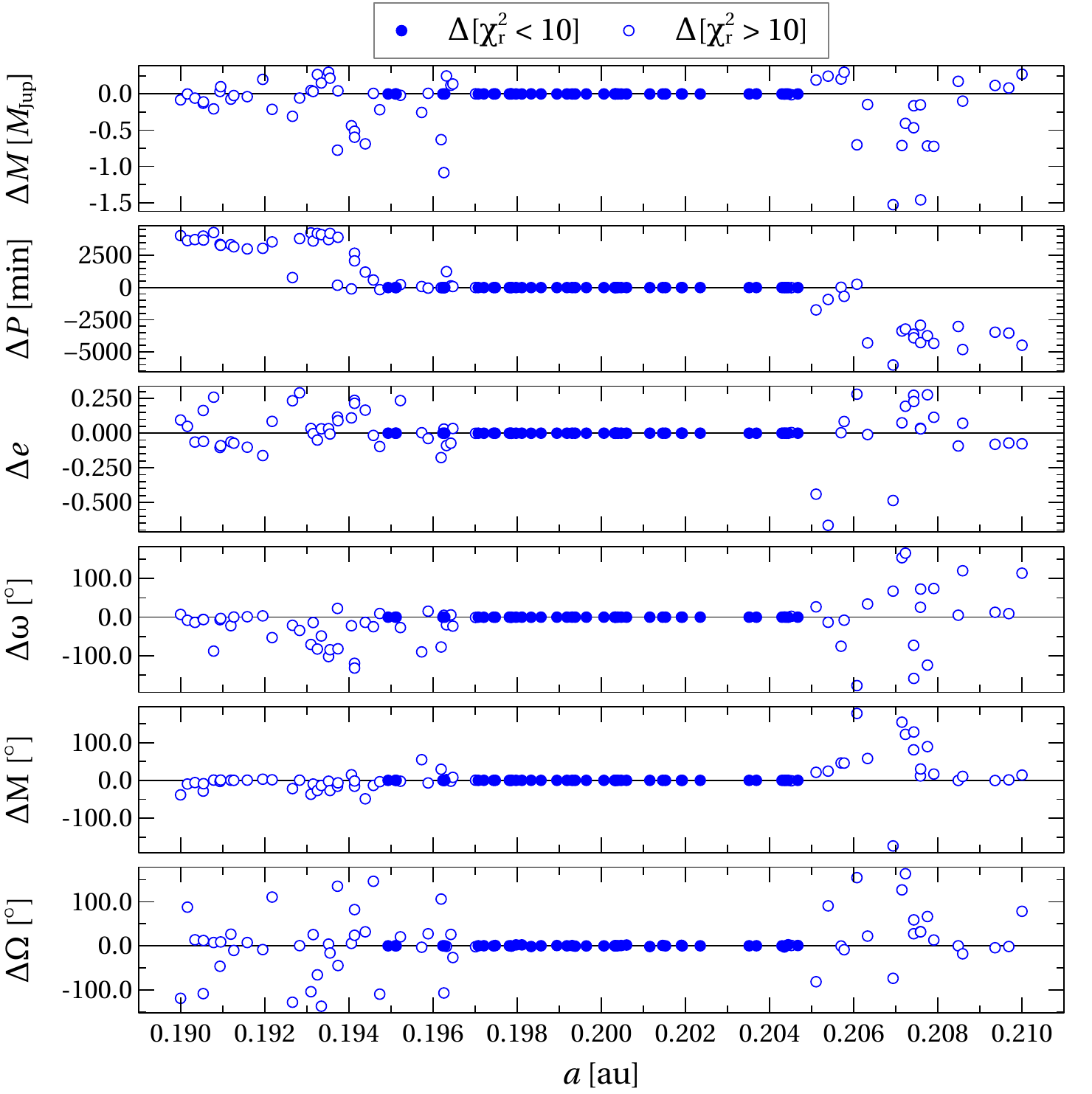}}
  \caption{
    Variations ($\Delta$) of the fitted parameters for different initial values of the semi-major axis of the planet c; each point corresponds to a different simulation.
	In this case, the input parameters (of planet c) are the parameters in Table~\ref{Tabsynth} used to generate the exact transit times.
	The goodness of the fit has been color-coded so that good fits ($\chi^2_\mathrm{r} < 10$) have been plotted as blue points and bad fits ($\chi^2_\mathrm{r} > 10$) as open circles.
	The small gaps are due to the random sampling used to generate the grid in the semi-major axis $a$.
  }
  \label{FigGRa2}
\end{figure}

To further test the robustness of the algorithm, we took the $T_{0,\mathrm{true}}$s (without added noise) and varied the initial semi-major axis of planet c from 0.19~AU to 0.21~AU with the \texttt{TRADES} grid+\texttt{LM} mode by fitting the same parameters of the previous test.
The algorithm nicely converged to the values from which the synthetic data were generated, except for initial parameters that are too far from the right solution.
  This is due to the known limitation of the \texttt{LM} algorithm, which converges to close local minima from an initial set of parameters.
Figure~\ref{FigGRa2} shows the variation of the parameter differences ($\Delta$) as function of the initial semi-major axis.
This test shows how well \texttt{TRADES} recovers the parameters in the case of a bad guess of the initial parameters.
\par

We measured the computational time required by \texttt{TRADES}, and we found that it can integrate (initial step size of 0.0001 day) a 3-body synthetic system for 3000 days writing the orbits, the Keplerian elements, and the constants of motion for each 0.1~days in about 2.3 seconds.
An integration of 1000 days has been performed in less than 1 second and in less than half second for 500 days of integration time, but most of the time have been spent writing files.
We want to stress that \texttt{TRADES} write these files only at the end of the simulations, so the real computation is faster than these estimates.
The time required by \texttt{TRADES} to complete the grid search was about 51 minutes with 10 processors of an Intel$^\circledR$ Xeon$^\circledR$ CPU E5-2680 based workstation.
For each combination of the initial parameters in the grid search, \texttt{TRADES} runs 10 times the \texttt{LM} to select the best value for the parameter \texttt{epsfcn}, that is needed to construct the initial Jacobian.
\par

We tested the \texttt{PSO+LM} algorithm by fitting the same parameters of the grid search with limited boundaries except for the semi-major axis of planet c, for which we used the same limit of the grid search ($a_\mathrm{c} = [0.19,0.21]$~AU).
We ran this test four times with 200 particles for 2000 iterations, and \texttt{TRADES} always returned the right parameter values in less than about 1 hour and 40 minutes with 10 processors.
\par

\section{Test case: Kepler-11 system}
\label{kepler-11}

The system \mbox{Kepler-11} \citep[KOI-157,][]{Lissauer2011Natur} has six transiting planets packed in less than $0.5$~AU, making a complex and challenging case to be tested with \texttt{TRADES}.\\
From the spectroscopic analysis of HIRES high-resolution spectra, \citet{Lissauer2011Natur} derived the stellar parameters (effective temperature, surface gravity, metallicity, and projected stellar equatorial rotation) and determined the mass and the radius of Kepler-11 star to be $0.95\pm0.10~M_{\sun}$ and $1.1\pm0.1~R_{\sun}$.\par

We first performed an analysis of the Kepler-11 system only on the data from the first three quarters of \textit{Kepler} observations published by \citet{Lissauer2011Natur} and supplementary information (SI).
We used the first circular model from the \citet[][SI]{Lissauer2011Natur} as an initial guess, which fixed the eccentricity and the longitude of the ascending node to zero for all the planets;
hereafter we call this model Lis2011 (see first column in Table~\ref{TabK11_best} for a summary of the orbital parameters).
We used this model because the authors did not provide any information about the mean anomalies (or the time of the passage at the pericenter) for those planets.
In this case, the argument of the pericenter, $\omega$, is undetermined, so we fixed it to $\omega=90\degr$ for each planet.
We then calculated the initial mean anomaly, $\mathrm{M}_{0}$ at the reference epoch $t_{\mathrm{epoch}}=2\,455\,190.0$ (BJD$_\mathrm{UTC}$\footnote{
	In the FITS header of \textit{Kepler} data the time standard is reported as Barycentric Julian Day in Barycentric Dynamical Time (BJD$_\mathrm{TDB}$),
	but it is specified in the KSCI-19059 Subsect.~3.4 that the correct time is Barycentric Julian Day in Coordinated Universal Time (BJD$_\mathrm{UTC}$).
	}
), setting the transit time \citep[$T_{0}$,][SI Table~S4]{Lissauer2011Natur} as the time of passage at the pericenter:
\begin{equation}
  \label{eq:mA1}
  \mathrm{M}_{0}=n \cdot (t_{\mathrm{epoch}} - T_{0}) \ ,
\end{equation}
where $n$ is the mean motion of the planet.\par

\begin{figure}
  \resizebox{\hsize}{!}{\includegraphics{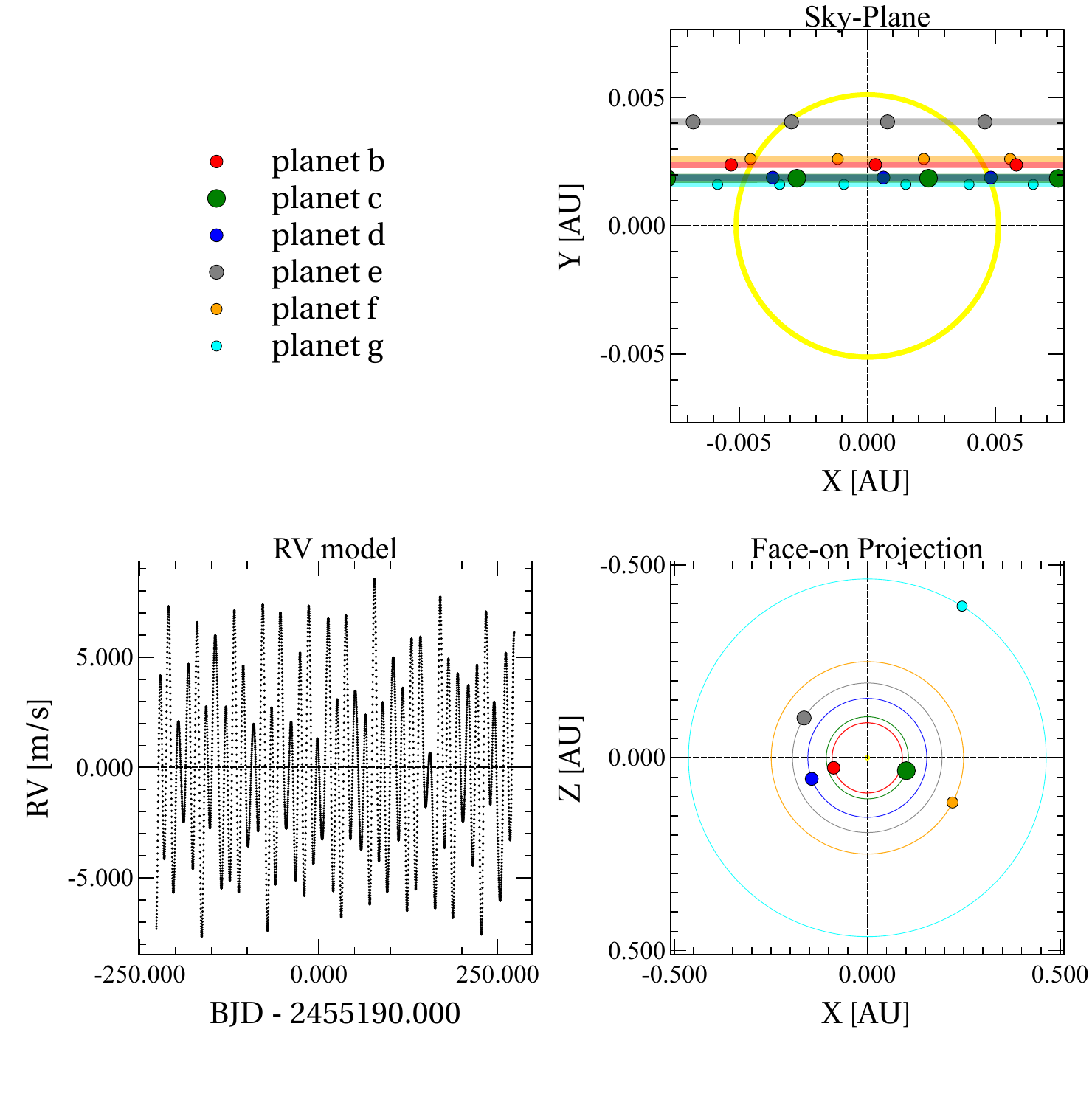}}
  \caption{
    Orbits of the Kepler-11 system with initial parameters from Lis2011 model (circular model, see Table~\ref{TabK11_best}).
    The planet marker size is scaled with the mass of the planet.
    \textit{Top-right}: `Sky Plane', Kepler-11 system as seen from the \textit{Kepler} satellite; we plotted only one orbit near a transit for each planet.
    Each circle is the position of a planet at a given integration time step.
    \textit{Bottom-right}: Projection of the system as seen face on. The big markers are the initial points of the integration.
    \textit{Bottom-left}: RV model from the simulation.
  }
  \label{FigK11orbit}
\end{figure}

\citet{Lissauer2011Natur} gave an upper limit of $300~M_{\oplus}$ on the mass of the planet Kepler-11g, while they set it to zero in the three dynamical models of the supplementary information , and we followed the same approach.
Figure~\ref{FigK11orbit} shows the orbits and RVs of the Kepler-11 planets, according to the Lis2011  model.\par

We fitted a linear ephemeris (Table~\ref{TabK-11_ephem}) to the observed transit times of each planet for the first three quarters and computed the Observed$-$Calculated ($O-C$) diagrams, where $O$ is $T_{0, \mathrm{obs}}$s and $C$ is the transit time calculated from the linear ephemeris.

We ran \texttt{TRADES} and fitted masses, periods, and  mean anomalies of each planet. Hereafter, the orbital solution we have determined with \texttt{TRADES} is named with a short ID of the system (K11 for Kepler-11 and K9 for Kepler-9) and a Roman number (K11-I, K11-II, and so on).
See solution K11-I in Table~\ref{TabK11_best} for a summary of the parameters determined with \texttt{TRADES} (with $2\sigma$ confidence intervals from bootstrap analysis), which agree with those published by \citet{Lissauer2011Natur}.
For each bootstrap run, we run 1000 iterations to obtain the confidence intervals at the $97.72$ percentile ($2\sigma$) of the distribution of each parameter.
We calculated the residuals as the difference between the observed and simulated $T_{0}$s.
The residuals after the \texttt{TRADES-LM} fit are smaller than those with simulated transit times, that obtained with original parameters, and the final $\chi^{2}_\mathrm{r}$ is around $\approx 1.25$ for 88 degrees of freedom (dof, calculated as the difference between the number of data and the number of fitted parameters).
\par

We also used the same initial conditions of solution K11-I, but this time we fitted the eccentricity and the argument of pericenter of all the planets.
In this case, the \texttt{LM} did not move from the initial conditions even if it properly ended the simulation and returned reasonable errors.
In the user guide of \texttt{MINPACK} \citep{MINPACK-1}, the user is warned to carefully analyze the case in which one has a null initial parameter.
We set the initial eccentricities to a small but non-zero value of $0.0001$.
This small change was able to let the \texttt{LM} algorithm to properly return reasonable parameter values; see solution K11-II in Table~\ref{TabK11_best} for a summary of the parameters.
\par

The resulting masses of the solutions K11-I and K11-II all agree within $2\sigma$ with the discovery paper \citep{Lissauer2011Natur} and with all the best-fit solutions determined by \citet{Migas2012}.
In the latter work, the authors presented different sets of orbital parameters determined with an approach similar to ours (direct $N$-body simulation with genetic algorithm, Levenberg-Marquardt, and bootstrap), but they directly fit the flux of \textit{Kepler} light curves (so-called dynamical photometric model) without fitting the transit times.

\subsection{Transit time analysis of the twelve quarters}
\label{K11_ext12}

Recently, \citet{Lissauer2013ApJ} analyzed the transit times covering fourteen quarters of \textit{Kepler} data (in long and short cadence mode).
A new independent extraction of $T_0$s from the light curves made by \citet[hereafter we call the dynamical model from this work Lis2013, see column five of Table~\ref{TabK11_best}]{Lissauer2013ApJ} led the authors to change the value of some parameters of the system. For example, they determined a mass of $2.9^{+2.9}_{-1.6}\ M_{\oplus}$ of the planet c that is lower than $15.82 \pm 2.21\ M_{\oplus}$ published in the discovery paper \citep{Lissauer2011Natur}.
Unfortunately the authors have not published the $T_0$, so we used the data from \citet{Mazeh2013arXiv} that recently published the transit times for twelve quarters of the \textit{Kepler} mission for 721 KOIs.

We used the linear ephemeris by \citet{Lissauer2013ApJ} to compute the $O-C$s for the $T_0$s from \citet{Mazeh2013arXiv}.
We found a remarkable mismatch with the $O-C$s plotted in the paper by \citet{Lissauer2013ApJ}.
We stress that the $T_{0}$s by \citet{Mazeh2013arXiv} are calculated with an automated algorithm.
It would be advisable to more carefully analyze the light curves determining the $T_0$s with higher precision, but it is not the purpose of this work.
We analyzed the system with all transit times from \citet{Mazeh2013arXiv} without any selection, and they lead to unphysical results.
We then decided to discard data with duration and depth of the transits that are $5\sigma$ away from the median values.
This selection defines the sample of $T_0$s for the first twelve quarters of Kepler-11 exoplanets on which we bases our next analysis.
\par

\begin{figure}[!]
  \resizebox{\hsize}{!}{\includegraphics[width=17cm]{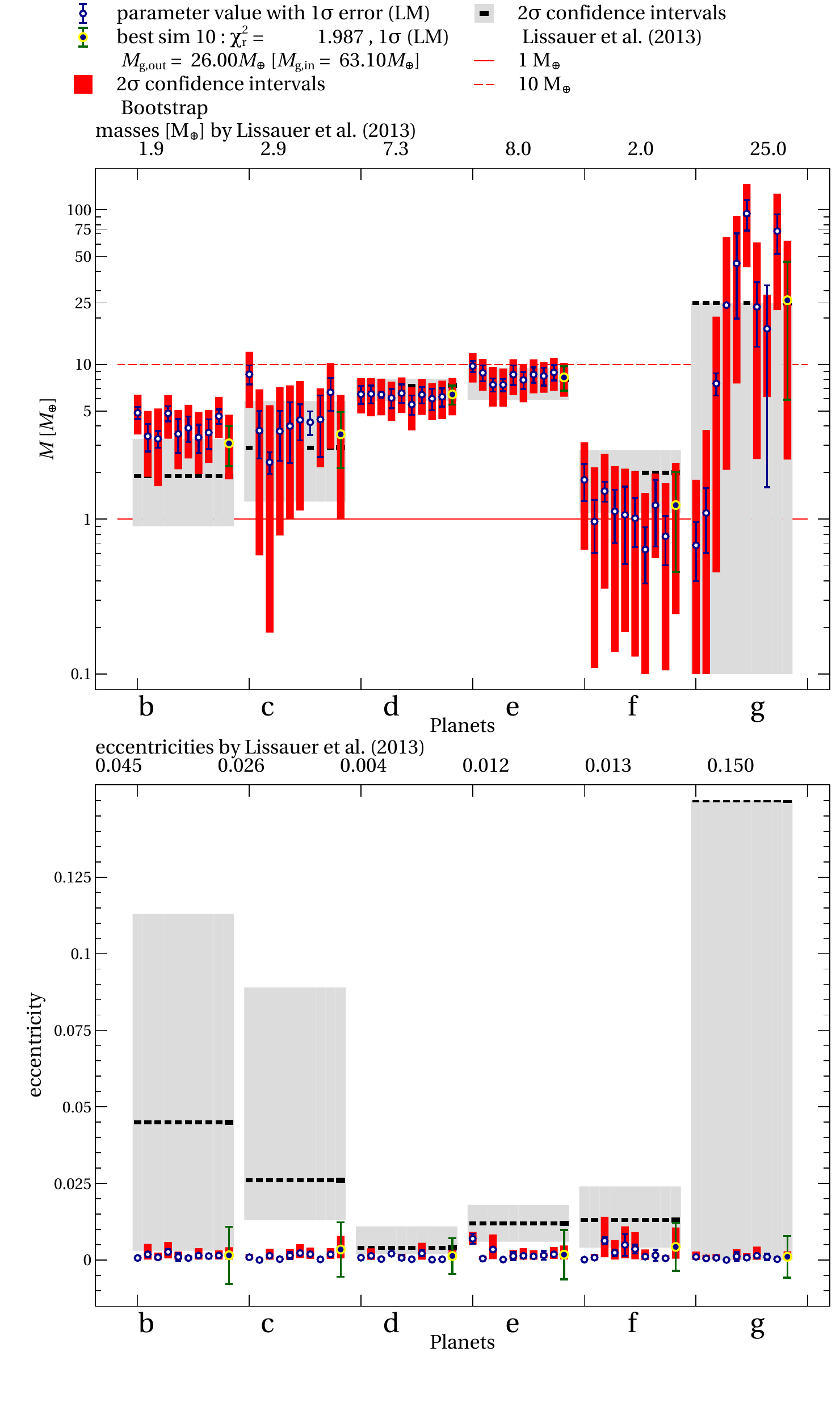}}
  \caption{
    Masses (\textit{upper-panel}) and eccentricities (\textit{lower-panel}) for the Kepler-11 planets,
    calculated with \texttt{TRADES} in \texttt{grid+LM} mode (white-blue circle with blue error bars, see the legend on top of the upper plot) and $2\sigma$ confidence intervals from bootstrap analysis (red filled bars), with initial eccentricity of 0.001 for each planet.
    The blue-yellow circle (dark-green error bars) is the best simulation (number 11, $\chi^2_\mathrm{r}$, calculated mass, $M_\mathrm{g,out}$, and input mass, $M_\mathrm{g,out}$, are reported in the legend at the top of the plots).
    The different simulations (different initial mass of planet g, $m_{\mathrm{g,in}}$) have been plotted from left (first simulation) to right (eleventh simulation) for each planet.
    Masses and eccentricities by \citet{Lissauer2013ApJ} plotted as black lines (values on top of the plots) with the $2\sigma$ confidence intervals (light-gray filled bars).
    Red lines at $1\ M_{\oplus}$ (solid) and at $10\ M_{\oplus}$ (dashed).
  }
  \label{FigK11_m_e_1}
\end{figure}

\begin{figure}[!]
	\resizebox{\hsize}{!}{\includegraphics{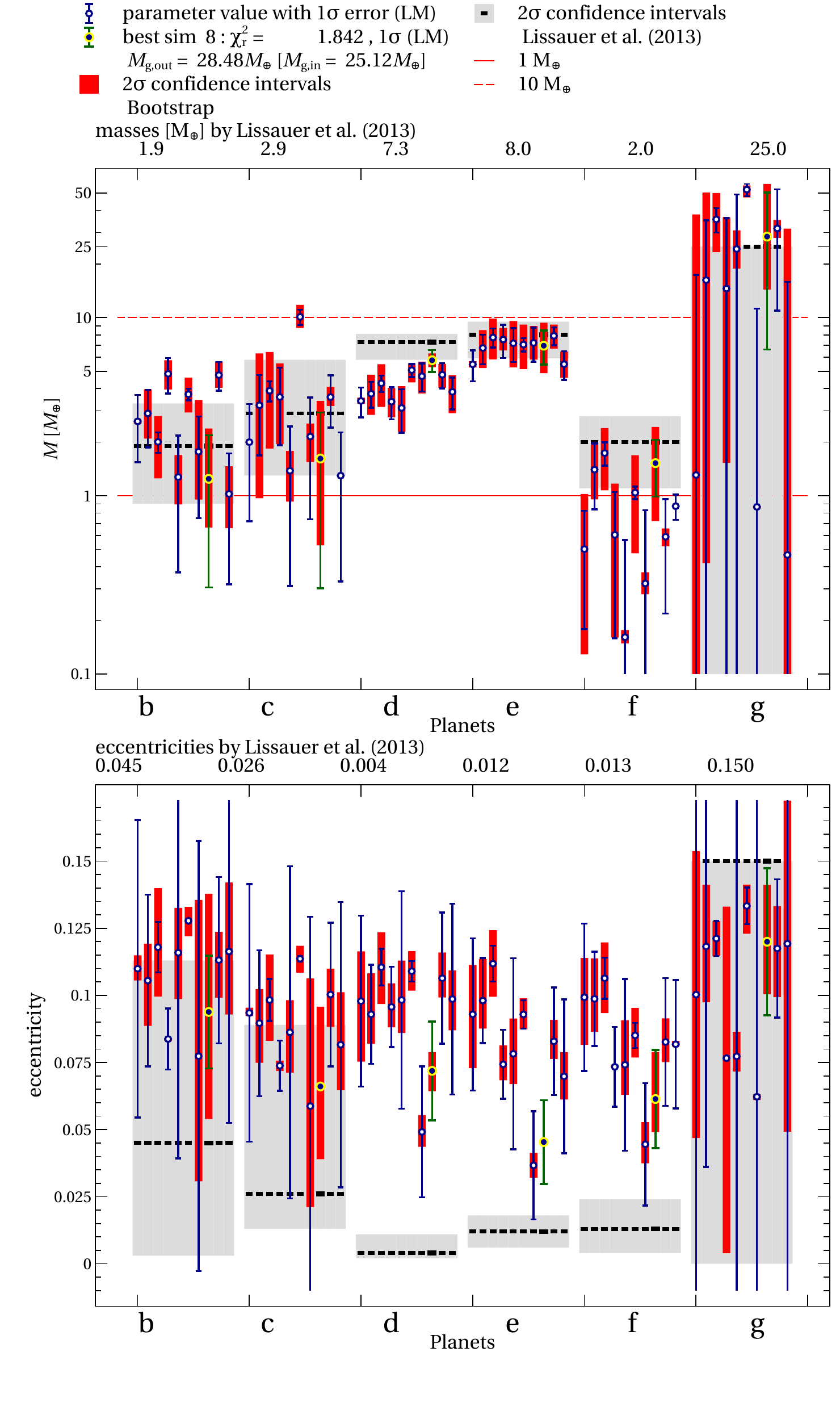}}
	\caption{
		Same plot as in Fig.~\ref{FigK11_m_e_1} but for simulations with initial eccentricities of 0.1.
	}
	\label{FigK11_m_e_2}
\end{figure}

\begin{figure}[!]
  \resizebox{\hsize}{!}{\includegraphics{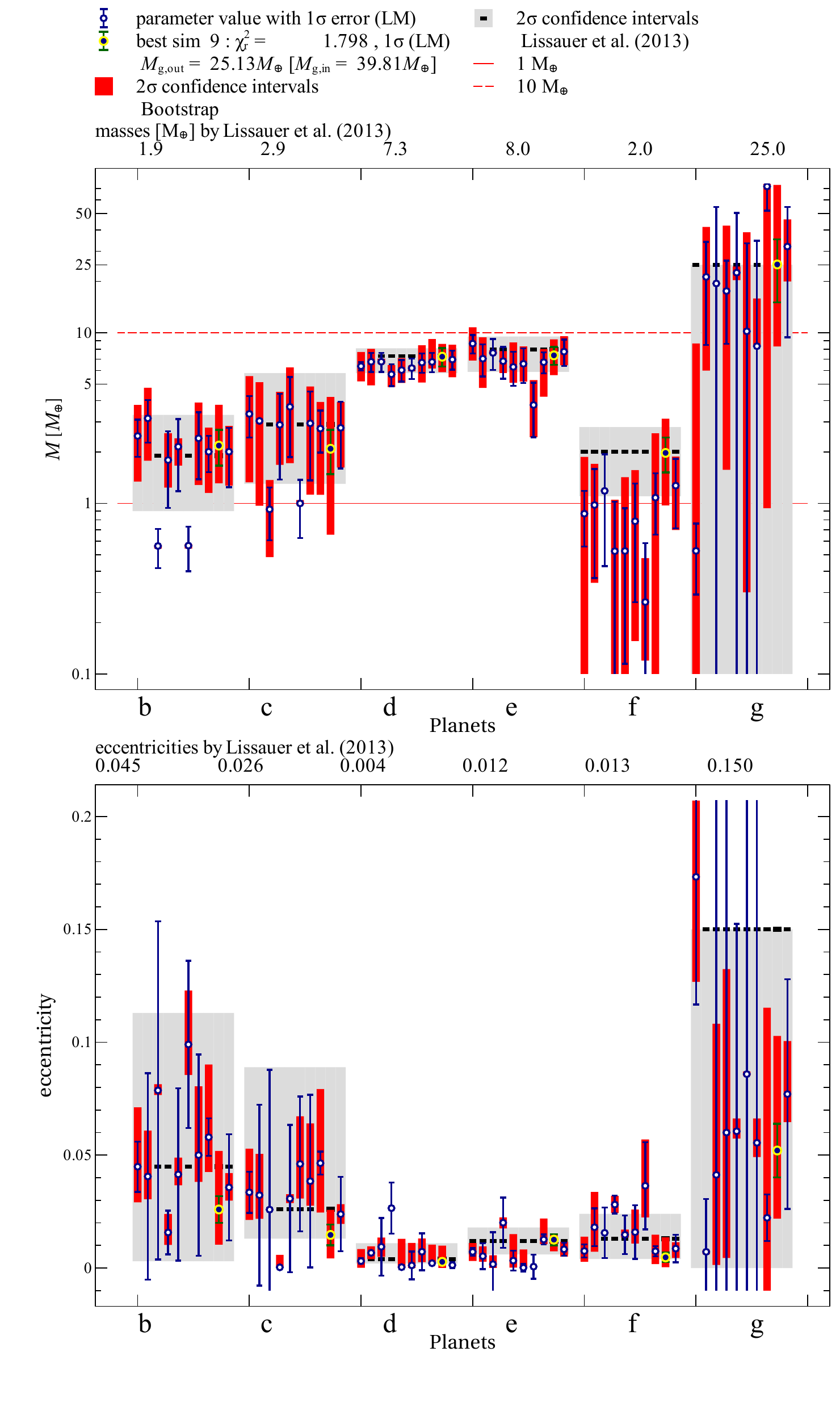}}
  \caption{
    Same plot as in Fig.~\ref{FigK11_m_e_1} but for simulations with different initial eccentricities:
    $e_\mathrm{b} = 0.05$, $e_\mathrm{c} = 0.05$, $e_\mathrm{d} = 0.001$, $e_\mathrm{e} = 0.005$, $e_\mathrm{f} = 0.005$, and $e_\mathrm{g} = 0.1$.
    The best solution of this plot is the so-called K11-III solution; see Table~\ref{TabK11_best} for the summary of the final parameters.
  }
  \label{FigK11_m_e_3}
\end{figure}

We ran simulations with \texttt{TRADES} in grid\texttt{+LM} mode on twelve quarters with initial set of parameters as in K11-II;
we fitted $M, P, e, \omega$, and  $\mathrm{M}$ of each planet ($\Omega = 0\degr$ fixed for all the planets).
In particular, we varied the mass of planet g from $1 \ M_{\oplus}$ to $100 \ M_{\oplus}$ with a logarithmic step (ten simulations including the boundaries of 1 and $100\ M_{\oplus}$) in the grid.
We repeated this set of simulations for three different initial values of the eccentricity:
in the first sample, we set the initial eccentricity of all planets to $0.001$;
in the second sample this is equal to $0.1$;
and in the third sample, we used a different value of the eccentricity for each planet,
which closer to the \citet{Lissauer2013ApJ} ones:
$e_\mathrm{b} = 0.05$, $e_\mathrm{c} = 0.05$, $e_\mathrm{d} = 0.001$, $e_\mathrm{e} = 0.005$, $e_\mathrm{f} = 0.005$, and $e_\mathrm{g} = 0.1$.
With these simulations, we intended to test whether a forest of local minima are met during the search for the lowest $\chi^2$ (see Figs.~\ref{FigK11_m_e_1}, \ref{FigK11_m_e_2}, and \ref{FigK11_m_e_3}).
According to Figs.~\ref{FigK11_m_e_1} and \ref{FigK11_m_e_2}, this indeed seems to be the case significantly complicating the identification of the real minimum.
The \texttt{LM} was not able to properly change the eccentricity of the planets that,  which got stuck close to the initial value in the majority of the cases.
Furthermore, when the initial eccentricity have been set to 0.1 the masses of planets d and f have decreased (Fig.~\ref{FigK11_m_e_2}) compared to those in the previous simulation (Fig.~\ref{FigK11_m_e_1}).
Maybe, this could be an effect due to the particular sample of $T_0$ we used, so it would be interesting to re-estimate the $T_0$ from the light curves and re-analyze the system.
\par

\begin{figure}[!]
	\resizebox{\hsize}{!}{\includegraphics{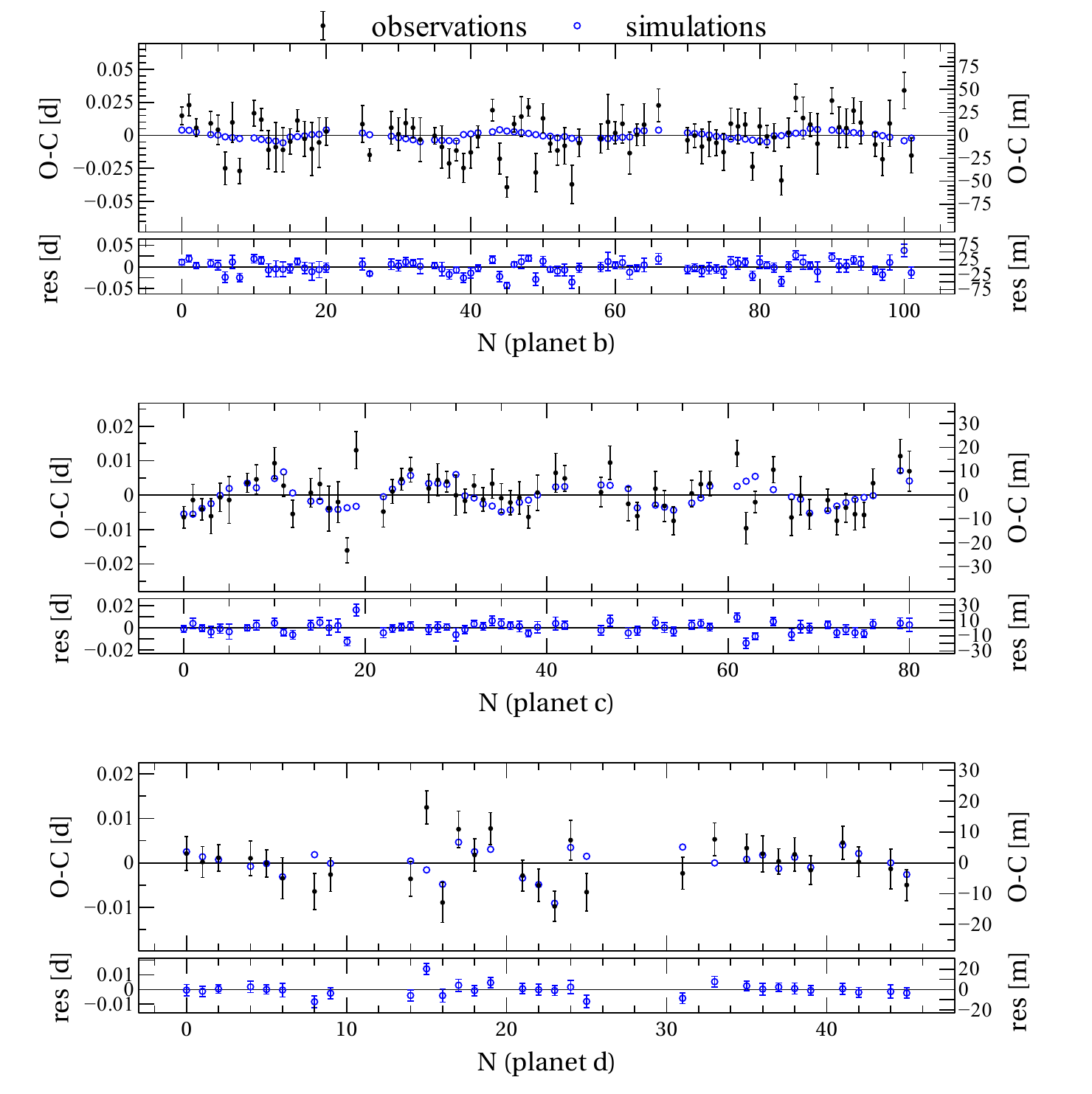}}
	\caption{
		$O-C$ diagrams for planets b, c, and d (from top to bottom) of the Kepler-11 system;
		the black filled circles are the observed points fitted by a linear ephemeris, the blue open circles are the simulated points fitted by the same linear ephemeris of the observations.
		The simulated points are calculated from the solution K11-III (best simulation in Fig.~\ref{FigK11_m_e_3}) in Table~\ref{TabK11_best}.
		Residual plots, as the difference between observed and simulated central time ($T_{0,\mathrm{obs}} - T_{0,\mathrm{sim}}$), are in the lower panel of each $O-C$ plot.
		The unit of measurement of the left $O-C$ y-axis is days (d) and minutes (m) are on the right.
		The N in the abscissa identifies the transit number respect to the reference transit time of the ephemeris of each body (second column of Table~\ref{TabK-11_ephem})
	}
	\label{FigK11_best_bcd}
\end{figure}

\begin{figure}[!]
	\resizebox{\hsize}{!}{\includegraphics{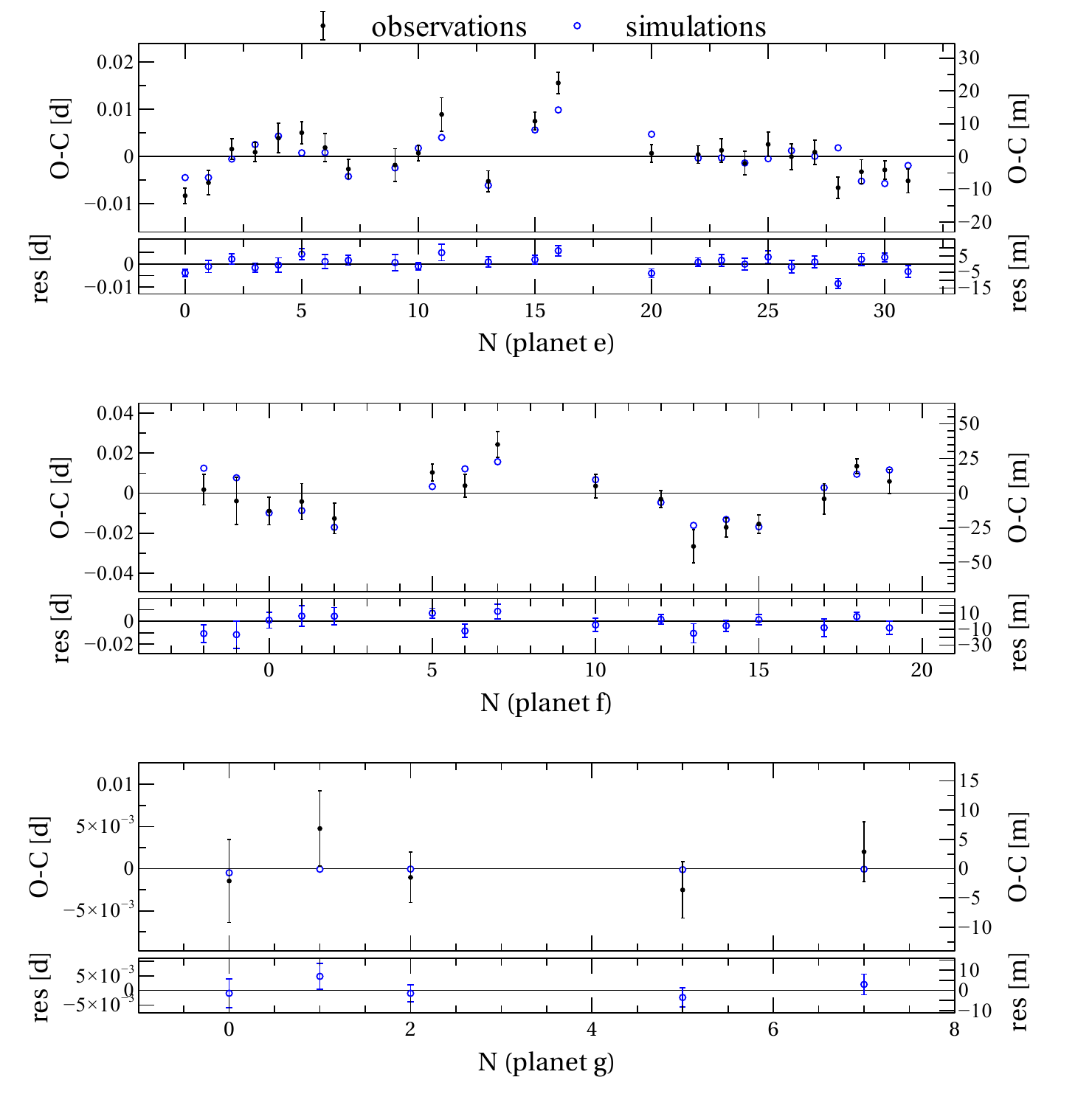}}
	\caption{
		Same as Fig.~\ref{FigK11_best_bcd} but for Kepler-11 planets e, f, and g (top to bottom).
	}
	\label{FigK11_best_efg}
\end{figure}

However, all the simulations have a final $\chi^2_\mathrm{r}$ of about 2 or lower;
the ninth simulation of the third set is our best solution (hereafter K11-III, see Table~\ref{TabK11_best}) with a $\chi^2_\mathrm{r} = 1.8$ (Fig.~\ref{FigK11_m_e_3}).
The resulting $O-C$ diagrams of the best simulation are shown in Figs.~\ref{FigK11_best_bcd} (planets b, c, and d) and \ref{FigK11_best_efg} (planet e, f, and g).
All the masses and the eccentricities of solution K11-III agree well with the values found by \citet{Lissauer2013ApJ}.
Some of our simulations converged to parameter values, which are different from those proposed by \citet{Lissauer2013ApJ}.
Furthermore, some simulations show very narrow confidence intervals.
This could be due both to the high complexity of the problem and to a strong selection effect: the distribution of the parameters in the bootstrap analysis are strongly bounded to the parameter values found by the \texttt{LM} algorithm.
\par

In Table~\ref{TabK-11_diff}, we report a brief summary of the main differences of the characteristics of the analysis that led us to each solution for the Kepler-11 system.\par

\begin{table*}[h]
  \caption{
    Parameters of the Kepler-11 system.
    Epoch of reference:\,\mbox{2\,455\,190.0 (BJD$_\mathrm{UTC}$)}.
  }
  \label{TabK11_best}
  \begin{center}
    \scriptsize
    \begin{tabular}{l r r r r r}
      \hline\hline\noalign{\smallskip}
      Parameter & Lis2011\phantom{}\tablefootmark{a} & K11-I\phantom{}\tablefootmark{b} & K11-II\phantom{}\tablefootmark{c} & Lis2013\phantom{}\tablefootmark{d} & K11-III\phantom{}\tablefootmark{e} \\

      \noalign{\smallskip}\hline\noalign{\smallskip}
      $ M_\mathrm{b}[M_\oplus] $ & $5.06\pm0.95$ & $ 5.51^{+ 1.91}_{- 2.04} \pm 1.15 $ & $ 5.03^{+2.29}_{-2.42} \pm 1.50 $ & $ 1.9^{+1.4}_{-1.0}$ & $ 2.18^{+1.60}_{-0.87} \pm 0.52 $ \\
      \noalign{\smallskip}
      $ R_\mathrm{b}[R_\oplus] $ & $1.97\pm0.19$ & & & $1.80^{+0.03}_{-0.05}$ & \\
      \noalign{\smallskip}
      $ P_\mathrm{b}[\mathrm{days}] $ & $10.3045\pm0.0003$ & $ 10.30459^{+ 0.00064}_{- 0.00060} \pm 0.00035 $ & $ 10.30446^{+0.00074}_{-0.00066} \pm 0.00070 $ & $10.3039^{+0.006}_{-0.0011}$ & $ 10.30448^{+0.00033}_{-0.00032} \pm 0.00019 $ \\
      \noalign{\smallskip}
      $ e_\mathrm{b}  $ & $0$ & & $ 0.00018^{+0.00094}_{-0.00017} \pm 0.00494 $ & $0.045^{+0.068}_{-0.042}$ & $ 0.026^{+0.026}_{-0.016} \pm 0.006 $ \\
      \noalign{\smallskip}
      $ \omega_\mathrm{b}[\degr] $ & $90$ & & $ 89.98^{+110.02}_{-112.11} \pm 220.29 $ & $45.00^{+101.31}_{-43.34}$ & $ 71.46^{+2.68}_{-2.63} \pm 17.06 $ \\
      \noalign{\smallskip}
      $ \mathrm{M}_\mathrm{b}[\degr] $ & $73.467 \pm 0.003$ & $ 73.46^{+ 0.17}_{- 0.15} \pm 0.09 $ & $ 73.48^{111.01}_{-110.73} \pm 219.60 $ & -- & $ 91.44^{+2.28}_{-2.41} \pm 16.19 $ \\
      \noalign{\smallskip}
      $ i_\mathrm{b}[\degr] $ & $88.5^{+1.0}_{-0.6}$ & & & $89.64^{+0.36}_{-0.18}$ & \\

      \noalign{\smallskip}\hline\noalign{\smallskip}
      $ M_\mathrm{c}[M_\oplus] $ & $15.82\pm2.21$ & $ 16.11^{+ 3.66}_{- 4.63} \pm 2.30 $ & $ 15.83^{+4.71}_{-4.94} \pm 3.99 $ & $2.9^{+2.9}_{-1.6}$ & $ 2.09^{+2.11}_{-1.43} \pm 0.61 $ \\
      \noalign{\smallskip}
      $ R_\mathrm{c}[R_\oplus] $ & $3.15\pm0.30$ & & & $2.87^{+0.05}_{-0.06}$ & \\
      \noalign{\smallskip}
      $ P_\mathrm{c}[\mathrm{days}] $ & $13.0247\pm0.0003$ & $ 13.02406^{+ 0.00041}_{- 0.00046} \pm 0.00026 $ & $ 13.02419^{+0.00045}_{-0.00054} \pm 0.00038 $ & $13.0241^{+0.0013}_{-0.0008}$ & $ 13.02426^{+0.00053}_{-0.00058} \pm 0.00028 $ \\
      \noalign{\smallskip}
      $ e_\mathrm{c}  $ & $0$ & & $ 0.00005^{+0.0006}_{-0.00005} \pm 0.00375 $ & $0.026^{+0.063}_{-0.013}$ & $ 0.015^{+0.011}_{-0.010} \pm 0.005 $ \\
      \noalign{\smallskip}
      $ \omega_\mathrm{c}[\degr] $ & $90$ & & $ 90.00^{+29.61}_{-34.03} \pm 100.29 $ & $51.34^{+128.63}_{-231.00}$ & $ 96.43^{+0.36}_{-0.24} \pm 29.56 $ \\
      \noalign{\smallskip}
      $ \mathrm{M}_\mathrm{c}[\degr] $ & $288.267 \pm 0.005$ & $ 288.27^{+0.06}_{-0.06} \pm 0.04 $ & $ 288.26^{+32.04}_{-31.18} \pm 99.66 $ & -- & $ 281.99^{+0.26}_{-0.37} \pm 28.71 $ \\
      \noalign{\smallskip}
      $ i_\mathrm{c}[\degr] $ & $89.0^{+1.0}_{-0.6}$ & & & $89.59^{+0.41}_{-0.16}$ & \\

      \noalign{\smallskip}\hline\noalign{\smallskip}
      $ M_\mathrm{d}[M_\oplus] $ & $5.69\pm1.27$ & $  5.97^{+ 2.32}_{- 2.57} \pm 1.36 $  & $ 5.67^{+2.70}_{-2.66} \pm 1.55 $ & $7.3^{+0.8}_{-1.5}$ & $ 7.24^{+1.37}_{-1.36} \pm 0.89 $ \\
      \noalign{\smallskip}
      $ R_\mathrm{d}[R_\oplus] $ & $3.43\pm0.32$ & & & $3.12^{+0.06}_{-0.07}$ & \\
      \noalign{\smallskip}
      $ P_\mathrm{d}[\mathrm{days}] $ & $22.6849\pm0.0007$ & $ 22.68509^{+ 0.00169}_{- 0.00167} \pm 0.00096 $ & $ 22.68494^{+0.00173}_{-0.00202} \pm 0.00135 $ & $22.6845^{+0.0010}_{-0.0009}$ & $ 22.68440^{+0.00095}_{-0.00095} \pm 0.00055 $ \\
      \noalign{\smallskip}
      $ e_\mathrm{d}  $ & $0$ & & $ 0.0001^{+0.0001}_{-0.0001} \pm 0.0074 $ & $0.004^{+0.007}_{-0.002}$ & $ 0.003^{+0.007}_{-0.003} \pm 0.001 $ \\
      \noalign{\smallskip}
      $ \omega_\mathrm{d}[\degr] $ & $90$ & & $ 90.00^{+7.72}_{-7.70} \pm 77.47 $ & $146.31^{+33.69}_{-146.31}$ & $ 102.52^{+0.22}_{-0.51} \pm 24.37 $ \\
      \noalign{\smallskip}
      $ \mathrm{M}_\mathrm{d}[\degr] $ & $69.245 \pm 0.002$ & $ 69.25^{+ 0.03}_{- 0.04} \pm 0.02 $ & $ 69.25^{+7.48}_{-8.43} \pm 76.58 $ & -- & $ 56.77^{+0.23}_{-0.50} \pm 24.25 $ \\
      \noalign{\smallskip}
      $ i_\mathrm{d}[\degr] $ & $89.3^{+0.6}_{-0.4}$ & & & $89.67^{+0.13}_{-0.16}$ & \\

      \noalign{\smallskip}\hline\noalign{\smallskip}
      $ M_\mathrm{e}[M_\oplus] $ & $8.22\pm1.58$ & $  8.44^{+3.38}_{-3.49} \pm 1.74 $ & $ 8.26^{+3.25}_{-3.43} \pm 2.03 $ & $8.0^{+1.5}_{-2.1}$ & $7.37^{+1.78}_{-1.73} \pm 0.89 $ \\
      \noalign{\smallskip}
      $ R_\mathrm{e}[R_\oplus] $ & $4.52\pm0.43$ & & & $4.19^{+0.07}_{-0.09}$ & \\
      \noalign{\smallskip}
      $ P_\mathrm{e}[\mathrm{days}] $ & $32.0001\pm0.0008$ & $ 32.00102^{+0.00300}_{-0.00366} \pm 0.00189 $ & $ 32.00044^{+0.00342}_{-0.00377} \pm 0.00305 $ & $31.9996^{+0.0008}_{-0.0013}$ & $ 32.00413^{+0.00173}_{-0.00207} \pm 0.00122 $ \\
      \noalign{\smallskip}
      $ e_\mathrm{e}  $ & $0$ & & $ 0.0002^{+0.0004}_{-0.0002} \pm 0.0089 $ & $0.012^{+0.006}_{-0.006}$& $ 0.013^{+0.003}_{-0.005} \pm 0.003 $ \\
      \noalign{\smallskip}
      $ \omega_\mathrm{e}[\degr] $ & $90$ & & $ 90.00^{+5.82}_{-5.20} \pm 1.11 $ & $-131.63^{+29.54}_{-25.75}$ & $ 204.69^{+0.26}_{-0.36} \pm 3.22 $ \\
      \noalign{\smallskip}
      $ \mathrm{M}_\mathrm{e}[\degr] $ & $122.211 \pm 0.003$ & $ 122.21^{+ 0.02}_{- 0.02} \pm 0.01 $ & $ 122.21^{+5.01}_{-6.06} \pm 0.04 $ & -- & $ 8.86^{+0.23}_{-0.40} \pm 3.19 $ \\
      \noalign{\smallskip}
      $ i_\mathrm{e}[\degr] $ & $88.8^{+0.2}_{-0.2}$ & & & $88.89^{+0.02}_{-0.02}$ & \\

      \noalign{\smallskip}\hline\noalign{\smallskip}
      $ M_\mathrm{f}[M_\oplus] $ & $1.90\pm0.95$ & $  2.15^{+ 1.85}_{- 1.76} \pm 0.98 $ & $ 2.19^{+1.98}_{-1.94} \pm 1.23 $ & $2.0^{+0.8}_{-0.9}$  & $ 1.98^{+1.16}_{-1.00} \pm 0.46 $ \\
      \noalign{\smallskip}
      $ R_\mathrm{f}[R_\oplus] $ & $2.61\pm0.25$ & & & $2.49^{+0.04}_{-0.07}$ & \\
      \noalign{\smallskip}
      $ P_\mathrm{f}[\mathrm{days}] $ & $46.6908\pm0.0010$ & $ 46.70131^{+ 0.00455}_{- 0.00851} \pm 0.00304 $  & $ 46.70114^{+0.00641}_{-0.00627} \pm 0.00688 $ & $46.6887^{+0.0029}_{-0.0038}$ & $ 46.68707^{+0.00384}_{-0.00575} \pm 0.00143 $ \\
      \noalign{\smallskip}
      $ e_\mathrm{f} $ & $0$ & & $ 0.000003^{+0.000001}_{-0.000002} \pm 0.000242 $ & $0.013^{+0.011}_{-0.009}$ & $ 0.005^{+0.010}_{-0.004} \pm 0.002 $ \\
      \noalign{\smallskip}
      $ \omega_\mathrm{f}[\degr] $ & $90$ & & $ 90.00^{+4.12}_{-4.09} \pm 0.03 $ & $-24.44^{+38.48}_{-47.12}$ & $ 8.58^{+0.32}_{-0.65} \pm 3.41 $ \\
      \noalign{\smallskip}
      $ \mathrm{M}_\mathrm{f}[\degr] $ & $297.667 \pm 0.006$ & $ 297.68^{+ 0.02}_{- 0.06} \pm 0.02 $ & $ 297.67^{+4.05}_{-4.14} \pm 0.06 $ & -- & $ 18.53^{+0.27}_{-0.66} \pm 3.38 $ \\
      \noalign{\smallskip}
      $ i_\mathrm{f}[\degr] $ & $89.4^{+0.3}_{-0.2}$ & & & $89.47^{+0.04}_{-0.04}$ & \\

      \noalign{\smallskip}\hline\noalign{\smallskip}
      $ M_\mathrm{g}[M_\oplus] $ & $<300$ & $  0.00^{+ 62.19}_{- 0.00} \pm 0.21 $ & $ 0.70^{+0.66}_{-0.54} \pm 41.50 $ & $<25$ & $ 25.13^{+48.33}_{-16.83} \pm 10.07 $ \\
      \noalign{\smallskip}
      $ R_\mathrm{g}[R_\oplus] $ & $3.66\pm0.35$ & & & $3.33^{+0.06}_{-0.08}$ & \\
      \noalign{\smallskip}
      $ P_\mathrm{g}[\mathrm{days}] $ & $118.3808\pm0.0025$ & $ 118.39734^{+0.00907}_{-0.00959} \pm 0.00517 $ & $ 118.39766^{+0.01080}_{-0.01053} \pm 0.01505 $ & $118.3809^{+0.0012}_{-0.0010}$ & $ 118.38030^{+0.00361}_{-0.00309} \pm 0.00248 $ \\
      \noalign{\smallskip}
      $ e_\mathrm{g}  $ & $0$ & & $ 0.0029^{+0.0015}_{-0.0014} \pm 0.2974 $ & $<0.15$ & $ 0.052^{+0.051}_{-0.030} \pm 0.012 $ \\
      \noalign{\smallskip}
      $ \omega_\mathrm{g}[\degr] $ & $90$ & & $ 90.01^{+0.63}_{-0.72} \pm 0.05 $ & $34.51^{+145.41}_{-214.50}$ & $ 97.00^{+0.29}_{-0.17} \pm 30.41 $ \\
      \noalign{\smallskip}
      $ \mathrm{M}_\mathrm{g}[\degr] $ & $211.997 \pm 0.005$ & $ 212.01^{+ 0.02}_{- 0.02} \pm 0.01 $ & $ 212.00^{+0.71}_{-0.64} \pm 0.05$ & -- & $ 205.71^{+0.29}_{-0.18} \pm 27.38 $ \\
      \noalign{\smallskip}
      $ i_\mathrm{g}[\degr] $ & $89.8^{+0.2}_{-0.2}$ & & & $89.87^{+0.05}_{-0.06}$ & \\
      \noalign{\smallskip}\hline

      $\chi^2 / \mathrm{dof}$ & $110.34 / 89$ & $ 110.15 / 88$ & $110.74 / 76 $ & & $341.75/ 190$\\
      $\chi^2_\mathrm{r}$ & $1.24$ & $1.25$ & $1.46$ & & $1.80$\\

    \end{tabular}
  \end{center}
  \tablefoot{
	Masses ($M$), periods ($P$), eccentricities ($e$), argument of pericenters ($\omega$), and mean anomaly ($\mathrm{M}$) of the best-fit simulation with $2\sigma$ confidence intervals from bootstrap analysis and $\pm 1\sigma$ from \texttt{LM}.
  Inclinations ($i$) fixed to the Lis2011 model.}\\
  \tablefoottext{a}{
	Dynamical model as reported in \citet[][SI]{Lissauer2011Natur} with circular orbit for each planet, $e$ fixed to $0$, and $\omega$ fixed to $90\degr$ ($e\cos \omega$ and $e\sin \omega$ set to zero in the discovery paper).}\\
  \tablefoottext{b}{
	Orbital solution from the analysis of $T_0$s from \citet{Lissauer2011Natur} for the first three quarters of \textit{Kepler} data.
	Parameters fitted: $M$, $P$, and $\mathrm{M}$.
	$e$ fixed to $0$ and $\omega$ fixed to $90\degr$}\\
  \tablefoottext{c}{
	Orbital solution from the analysis of $T_0$s from \citet{Lissauer2011Natur} for the first three quarters of \textit{Kepler} data.
    Parameters fitted: $M$, $P$, $e$, $\omega$, and $\mathrm{M}$.}\\
  \tablefoottext{d}{
	Dynamical model from \citep{Lissauer2013ApJ}.
	Parameters determined from the analysis of 14 quarters of \textit{Kepler} data.
	The values of the mean anomaly were not reported in the paper (neither the time of passage at the pericenter).}\\
  \tablefoottext{e}{
	Best orbital solution (simulation number 9) of Fig.~\ref{FigK11_m_e_3} from the analysis of $T_0$s from \citet{Mazeh2013arXiv} for the first 12 quarters of \textit{Kepler} data.
    Parameters fitted: $M$, $P$, $e$, $\omega$, and $\mathrm{M}$.
    }
\end{table*}

\begin{table}
  \caption{
    Ephemeris fitted to the first three quarters of data of the Kepler-11 system.
  }
  \label{TabK-11_ephem}
  \centering
  \small
  \begin{tabular}{c c c}
    \hline\hline\noalign{\smallskip}
    Planet & $T_{0}$ [BJD$_\mathrm{UTC}$] & $P$ [days]\\
    \hline\noalign{\smallskip}
    b & $2\,455\,187.88389 \pm 0.00028$ & $ 10.30375 \pm 0.00002$ \\
    c & $2\,455\,205.62519 \pm 0.00014$ & $ 13.02502 \pm 0.00001$ \\
    d & $2\,455\,185.63958 \pm 0.00027$ & $ 22.68718 \pm 0.00004$ \\
    e & $2\,455\,147.13846 \pm 0.00021$ & $ 31.99589 \pm 0.00006$ \\
    f & $2\,455\,151.40372 \pm 0.00089$ & $ 46.68877 \pm 0.00036$ \\
    g & $2\,455\,120.29008 \pm 0.00286$ & $118.37774 \pm 0.00237$ \\
    \noalign{\smallskip}\hline
  \end{tabular}
\end{table}

\begin{table*}
  \caption{
    Main differences in the Kepler-11 analysis for each solution.
  }
  \label{TabK-11_diff}
  \centering
  \small
  \begin{tabular}{l c c c}
    \hline\hline\noalign{\smallskip}
    & K11-I & K11-II & K11-III \\
    \hline\noalign{\smallskip}
    Quarters & 1 to 3 & 1 to 3 & 1 to 12 \\
    Initial parameters & Lis2011 \citep{Lissauer2011Natur} & K11-I & K11-II and $e$ from Lis2013 \citep{Lissauer2013ApJ} \\
    Number of fitted parameters & 18 & 30 & 30 \\
    Degrees of freedom (dof) & 88 & 76 & 190 \\
    \texttt{TRADES} mode & \texttt{LM} & \texttt{LM} & \texttt{grid} ($M_\mathrm{g}$) $+$ \texttt{LM}\\
    Bootstrap & yes & yes & yes \\
    $\chi^2_\mathrm{r}$ & 1.25 & 1.46 & 1.80 \\
    \noalign{\smallskip}\hline
  \end{tabular}
\end{table*}

\section{Test case: Kepler-9 system}
\label{kepler-9}

Another ideal benchmark for testing \texttt{TRADES} is the multiple planet system Kepler-9 (KOI-377).
The star Kepler-9 is a Solar-like G2 dwarf with a magnitude $V=13.9$ \citep{Holman2010Sci}, mass of $1.07\pm0.05~M_{\sun}$, and radius $R_{\star}=1.02\pm0.05~R_{\sun}$ \citep{Torres2011BLENDER}.
From the first three quarters of the \textit{Kepler} data, \citet{Holman2010Sci} identified two transiting Saturn-sized planet candidates (Kepler-9 b and c with radii of about $\sim 0.8~R_{\mathrm{Jup}}$) near the 2:1 mean motion resonance (MMR).
They detected an additional signal related to a third, smaller planet (KOI-377.03, estimated radius $\sim 1.5\, R_{\oplus}$), which is validated with \texttt{BLENDER} in \citet{Torres2011BLENDER} but still unconfirmed.
The last planet is not input in our simulations, given that there is no confirmation by spectroscopic follow-up so far (the expected RV semi-amplitude of about $\sim 1.5$~ms$^{-1}$ would not increase the scatter in the RV data).
Moreover, \citet{Holman2010Sci} stated that the dynamical influences of the fourth body on other planets is undetectable on \textit{Kepler} data (TTV amplitude of the order of ten seconds).\\
From the analysis of the $\mathrm{d}P/\mathrm{d}t$ of the parabolic fit (quadratic ephemeris) for the $T_0$s of each planet \citet{Holman2010Sci} inferred the masses of Kepler-9b and Kepler-9c to be $0.252 \pm 0.013 \, M_{\mathrm{Jup}}$, and $0.171 \pm 0.013 \, M_{\mathrm{Jup}}$, respectively; they used the RV measurements from six spectra with the HIRES echelle spectrograph at Keck Observatory \citep{Vogt1994SPIE} only to put a constraint on the masses.
\citet{Holman2010Sci} set an upper limit to the mass of the KOI-377.03 of about $7\, M_{\oplus}$, but they could not fix the lower mass limit.
The authors proposed $1\, M_{\oplus}$ for a volatile-rich planet with a hot extended atmosphere.
\citet{Torres2011BLENDER} could not determine a mass value for KOI-377.03 but estimated a radius of $1.64^{+0.19}_{-0.14}\, R_{\oplus}$.\par

\begin{figure}[h]
  \resizebox{\hsize}{!}{\includegraphics{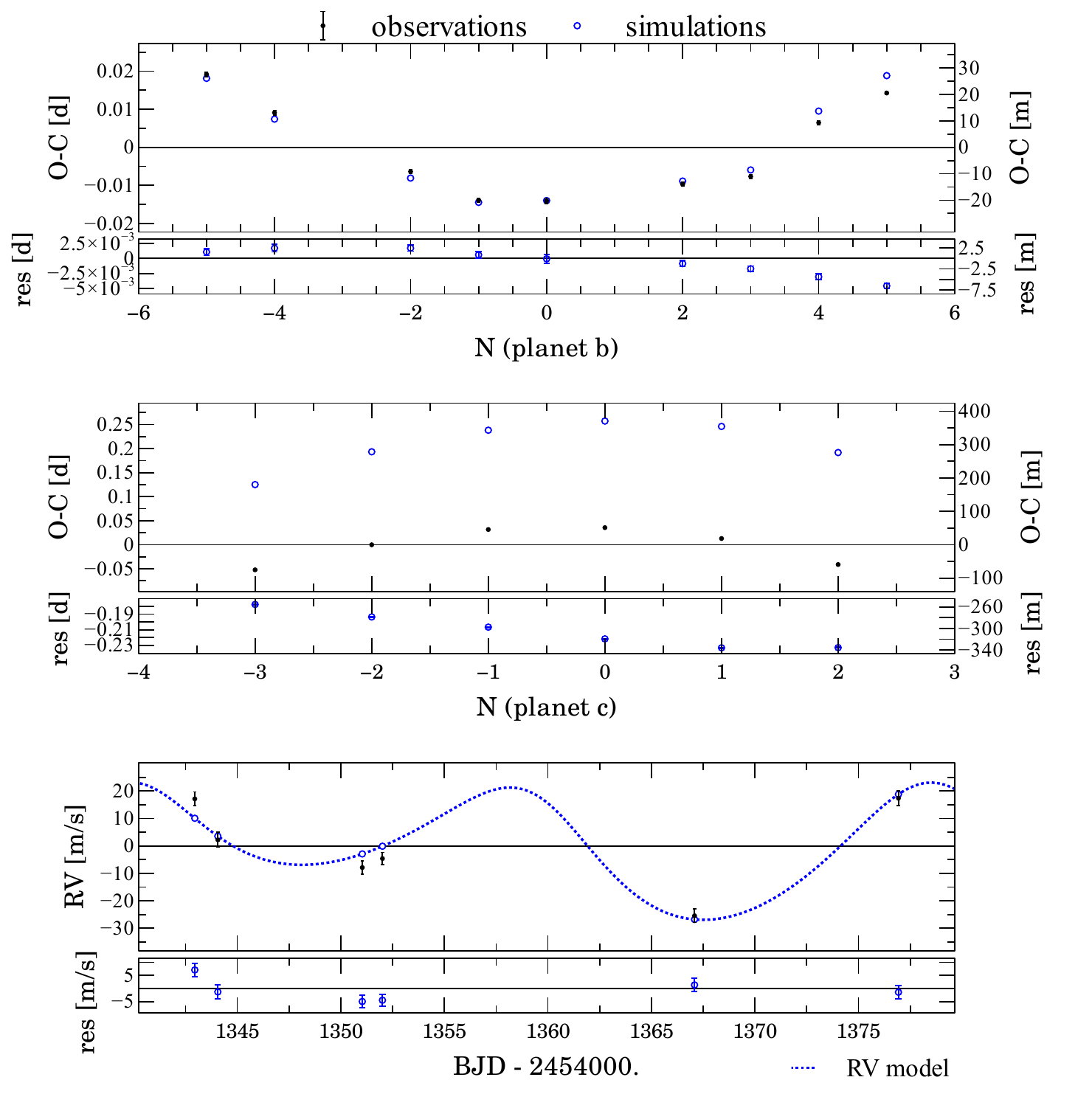}}
  \caption{
    $O-C$ diagrams (with residuals) from linear ephemeris for planet Kepler-9b (top panel) and Kepler-9c (middle panel) with the discovery paper's parameters (see column two of Table~\ref{TabPar});
    observations plotted as solid black circles, simulations plotted as open blue circles.
    The parameter $\mathrm{N}$, in the abscissa, has the same meaning of Fig.~\ref{FigK11_best_bcd}.
    \textit{Bottom panel} shows the RV observations as solid black circles, simulations at the same BJD$_\mathrm{UTC}$ as open blue circles, and the dotted blue line is the RV model for the whole simulation.
  }
  \label{FigHolman}
\end{figure}

We assumed the orbital parameters of the two planets at $t_\mathrm{epoch} = 2\,455\,088.212 \, \mathrm{BJD}_\mathrm{UTC}$ from Table~S6 in \citet[][ supporting on-line material, SOM]{Holman2010Sci} and set $i_\mathrm{c}=89\fdg12$ (Holman 2012, priv.~comm.; the value of $88\fdg12$ reported in the supporting on-line material is inconsistent with the transit geometry).
We simulated the system with \texttt{TRADES} without fitting any parameter, spanning the first three quarters of the \textit{Kepler} observations.
We fitted a linear ephemeris (see Table~\ref{TabK-9eph}) to the observations and compared the resulting $O-C$ diagrams with those from the simulations (Fig.~\ref{FigHolman}).
With the parameters from \citet{Holman2010Sci}, we obtained a simulated $O-C$ for Kepler-9c, which is systematically offset from the observed data points by $\sim300$ minutes (see Fig.~\ref{FigHolman}, middle panel).
In the bottom panel of the Fig.~\ref{FigHolman}, we plot the RV model compared to the observations and the residuals.\\

We also reported the quadratic ephemeris for comparison with the discovery paper in Table~\ref{TabK-9eph}.
Our linear and quadratic ephemeris in Table~\ref{TabK-9eph} adopt the transit closest to the median epoch as time of reference for each body, while \citet{Holman2010Sci} used the last transit time as reference for Kepler-9c.
\par

\begin{table}[h]
  \caption{
    Linear and quadratic ephemeris (in BJD$_\mathrm{UTC}$) fitted to data of Kepler-9 system.
  }
  \label{TabK-9eph}
  \centering
  \small
  \begin{tabular}{c c}
    \hline\hline\noalign{\smallskip}
    ephemeris & Kepler-9b\\
    \noalign{\smallskip}\hline\noalign{\smallskip}
    linear & $2\,455\,073.448177 \pm 0.000069$ \\
    & $19.243719 \pm 0.000020$ \\
    \noalign{\smallskip}\hline\noalign{\smallskip}
    quadratic & $2\,455\,073.433861 \pm 0.012302$ \\
    & $19.243164 \pm 0.002317$ \\
    & $0.001271 \pm 0.000846$ \\
    \noalign{\smallskip}\hline\noalign{\smallskip}
  \end{tabular}
  \begin{tabular}{c c}
    \noalign{\smallskip}\hline\hline\noalign{\smallskip}
    & Kepler-9c\\
    \noalign{\smallskip}\hline\noalign{\smallskip}
    linear & $2\,455\,086.276884 \pm 0.000121$ \\
    & $38.972972 \pm 0.000072$ \\
    \noalign{\smallskip}\hline\noalign{\smallskip}
    quadratic \phantom{}\tablefootmark{a} & $2\,455\,086.311873 \pm 0.014707$ \\
    & $38.962410 \pm 0.006893$ \\
    & $-0.013413 \pm 0.004030$ \\
    \noalign{\smallskip}\hline\noalign{\smallskip}
  \end{tabular}
  \tablefoot{
    \tablefoottext{a}{We used the central transit time as reference for the quadratic fitting, while \citet{Holman2010Sci} used the last time.}
  }
\end{table}

To investigate whether this behavior is due to bugs in \texttt{TRADES}, we ran a second analysis with the \texttt{MERCURY} package \citep{Chambers1997MERCURY}.
We simulated the same system with \texttt{MERCURY} and used the same technique as described in Sect.~\ref{Tdet} to calculate the central time of the simulated transits.
The maximum absolute difference between the mid-transit times from \texttt{TRADES} and \texttt{MERCURY} (with RADAU15 and Hybrid integrator) is $\sim0.16$ seconds for an integration of 500 days.
We did the calculation of the Keplerian orbital elements both for \texttt{TRADES} and \texttt{MERCURY}, and we verified the trend of the $X$ coordinate (the coordinate used as \textit{alarm} in case of eclipses for Kepler-9) of each planet as function of time (in a range of time around an observed transit):
we did not find any difference or unexpected behavior between \texttt{TRADES} and \texttt{MERCURY}.
These tests support our results showing that the problem is not in the integrator or in the subroutine used to calculate the transit times.
\par

\begin{figure}[h]
  \resizebox{\hsize}{!}{\includegraphics{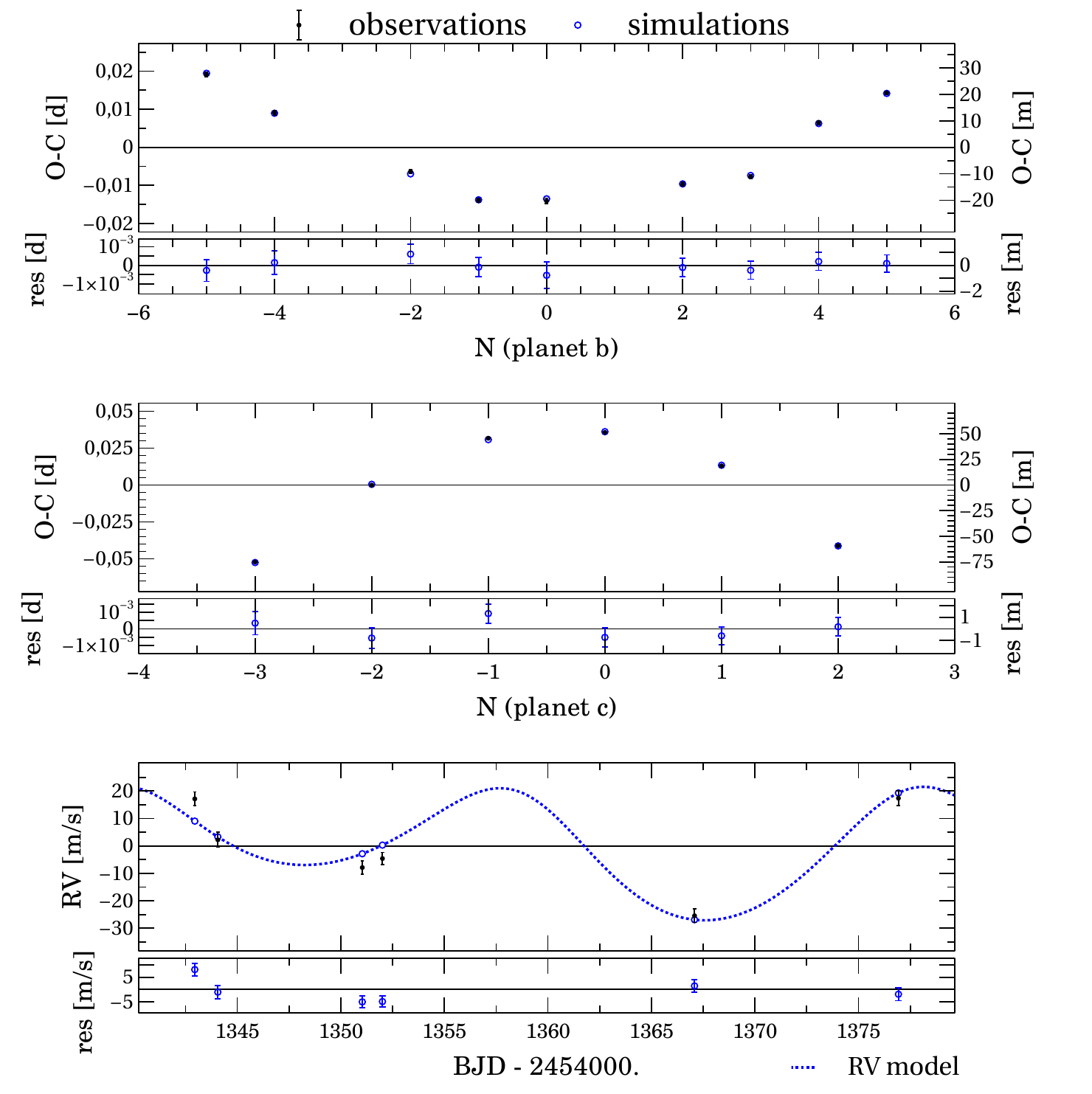}}
  \caption{
    Kepler-9 system: same plots as in Fig.~\ref{FigHolman} but with the parameters determined with \texttt{TRADES-LM} (K9-I of Table~\ref{TabPar}).
  }
  \label{FigLMfit11par}
\end{figure}

Then, we fitted $M$, $P$, $e$, $\omega$, and M (mean anomaly) of both planets and $\Omega$ of planet c using the \texttt{LM} algorithm in \texttt{TRADES}.
We found that the new values are consistent for all the fitted parameters with those by \citet[][see column one and two of Table~\ref{TabPar} for a comparison]{Holman2010Sci}.
Only one parameter, $P_\mathrm{c}$, agrees with the discovery paper within $2 \sigma$.
The small changes in the parameter values are enough to explain the $O-C$ offset of planet c.
The mean longitudes ($\lambda = \Omega + \omega + \mathrm{M}$) of the two planets differs from the two solution only by few degrees, but this determines a small misalignment of the initial condition that could have a strong effect in MMR configuration.
This simulation gives a $\chi^2 \approx 28.39$, for 10 dof, resulting in a $\chi^2_\mathrm{r} \approx 2.839$.
The results are summarized in Table~\ref{TabPar} (solution K9-I) and in Fig.~\ref{FigLMfit11par} the $O-C$s and the RV diagrams are plotted (notations and colors as in Fig.~\ref{FigHolman}).
\par

\begin{table*}[h]
  \caption{
    Parameters of the Kepler-9 system at epoch \mbox{$t_\mathrm{epoch} = 2\,455\,088.212 \, \mathrm{BJD}_\mathrm{UTC}$}.
  }
  \label{TabPar}
  \centering
  \small
  \begin{tabular}{l r r r }
    \hline\hline\noalign{\smallskip}
    Parameter & \multicolumn{1}{c}{\citet{Holman2010Sci}\phantom{}\tablefootmark{a}} & \multicolumn{1}{c}{K9-I\phantom{}\tablefootmark{b}} & \multicolumn{1}{c}{K9-II\phantom{}\tablefootmark{c}}\\
    \hline\noalign{\smallskip}
    $M_{\star} [M_{\sun}]$ & $1.0\pm0.1$ & & \\
    $R_{\star} [R_{\sun}]$ & $1.1\pm0.09$ & & \\

    \noalign{\smallskip}\hline\noalign{\smallskip}

    $M_\mathrm{b} [M_\mathrm{Jup}]$ & $ 0.252\pm0.013$ & $ 0.246^{+0.008}_{-0.008} \pm 0.014 $ & $ 0.137^{+0.001}_{-0.001} \pm 0.002 $ \\
    $R_\mathrm{b} [R_\mathrm{Jup}]$ & $ 0.842\pm0.069$ & & \\
    $P_\mathrm{b} [\mathrm{days}]$ & $ 19.2372\pm0.0007$ & $ 19.23686^{+0.00041}_{-0.00032} \pm 0.00051 $ & $ 19.23876^{+0.00004}_{-0.00004} \pm 0.00006 $ \\
    \noalign{\smallskip}
    $e_\mathrm{b}$ & $0.151\pm0.034$ & $ 0.131^{+0.008}_{-0.006} \pm 0.016 $ & $ 0.058^{+0.001}_{-0.001} \pm 0.002 $ \\
    $i_\mathrm{b} [\degr]$ & $88.55\pm0.25$ & & \\
    $\omega_\mathrm{b} [\degr]$ & $18.56\pm13.69$ & $ 18.91^{+0.60}_{-0.92} \pm 14.58 $ & $ 356.06^{+0.11}_{-0.21} \pm 0.44 $ \\
    \noalign{\smallskip}
    $\mathrm{M}_\mathrm{b} [\degr]$ & $332.15\pm14.06$ & $ 333.79^{+0.89}_{-0.97} \pm 14.27 $ & $ 3.78^{+0.22}_{-0.20} \pm 0.60 $ \\
    $\Omega_\mathrm{b} [\degr]$ & $0$ (fixed) & & \\

    \noalign{\smallskip}\hline\noalign{\smallskip}

    $M_\mathrm{c} [M_\mathrm{Jup}]$ & $0.171\pm0.013$ & $ 0.169^{+0.005}_{-0.006} \pm 0.017 $ & $ 0.094^{+0.001}_{-0.001} \pm 0.002 $ \\
    $R_\mathrm{c} [R_\mathrm{Jup}]$ & $0.823\pm0.067$ & & \\
    $P_\mathrm{c} [\mathrm{days}]$ & $38.992\pm0.005$ & $ 38.97897^{+0.00182}_{-0.00222} \pm 0.00336 $ & $ 38.98610^{+0.00020}_{-0.00021} \pm 0.00043 $ \\
    \noalign{\smallskip}
    $e_\mathrm{c}$ & $0.133\pm0.039$ & $ 0.119^{+0.004}_{-0.003} \pm 0.012 $ & $ 0.068^{+0.001}_{-0.001} \pm 0.001 $ \\
    $i_\mathrm{c} [\degr]$ & $89.12\pm0.17$\tablefootmark{d} & & \\
    $\omega_\mathrm{c} [\degr]$ & $101.31\pm47.05$ & $ 102.85^{+0.43}_{-0.51} \pm 8.04 $ & $ 167.57^{+0.01}_{-0.01} \pm 0.01 $ \\
    \noalign{\smallskip}
    $\mathrm{M}_\mathrm{c} [\degr]$ & $6.89\pm47.20$ & $ 7.48^{+0.41}_{-0.35} \pm 6.10 $ & $ 307.43^{+0.06}_{-0.05} \pm 0.07 $ \\
    \noalign{\smallskip}
    $\Omega_\mathrm{c} [\degr]$ & $2\pm3$ & $ 1.63^{+0.07}_{-0.11} \pm 1.19 $ & $ 359.89^{+0.30}_{-0.98} \pm 0.02 $\\

    \noalign{\smallskip}\hline\noalign{\smallskip}

    $\chi^2/\mathrm{dof} $ & & $ 28.382 / 10 $ & $ 80.852 / 56 $ \\
    $\chi^2_\mathrm{r} $ & & $2.84 $ & $1.44$ \\

    \noalign{\smallskip}\hline

  \end{tabular}
  \tablefoot{Results for the analysis of the Kepler-9 system with \texttt{TRADES} using the masses, the period, the eccentricity, the argument of pericenter, the mean anomaly of both planets, and the longitude of node of Kepler-9c as fitting parameters.\\
    \tablefoottext{a}{Parameters from the SOM of \citet{Holman2010Sci}.}\\
    \tablefoottext{b}{Analysis of \texttt{TRADES} with initial parameters and $T_0$s from \citet[][ SOM]{Holman2010Sci}. Transits and RVs fit.}\\
    \tablefoottext{c}{Analysis with \texttt{TRADES+PSO+LM} and $T_0$s from \citet{Mazeh2013arXiv}.
      Initial parameter boundaries were large enough to contains both solutions K9-I and by \citet{Holman2010Sci}. We fit only the transit times, and ignored the six RV points.}\\
    \tablefoottext{d}{
      The authors confirmed a typo in the inclination of Kepler-9c in the SOM (Holman 2012, priv.~comm.), considering that the value of $88\fdg12$ reported is inconsistent with the transit geometry.}
  }
\end{table*}

\subsection{Transit time analysis of the twelve quarters}
\label{k9_ext12}

\begin{figure}[h]
  \resizebox{\hsize}{!}{\includegraphics{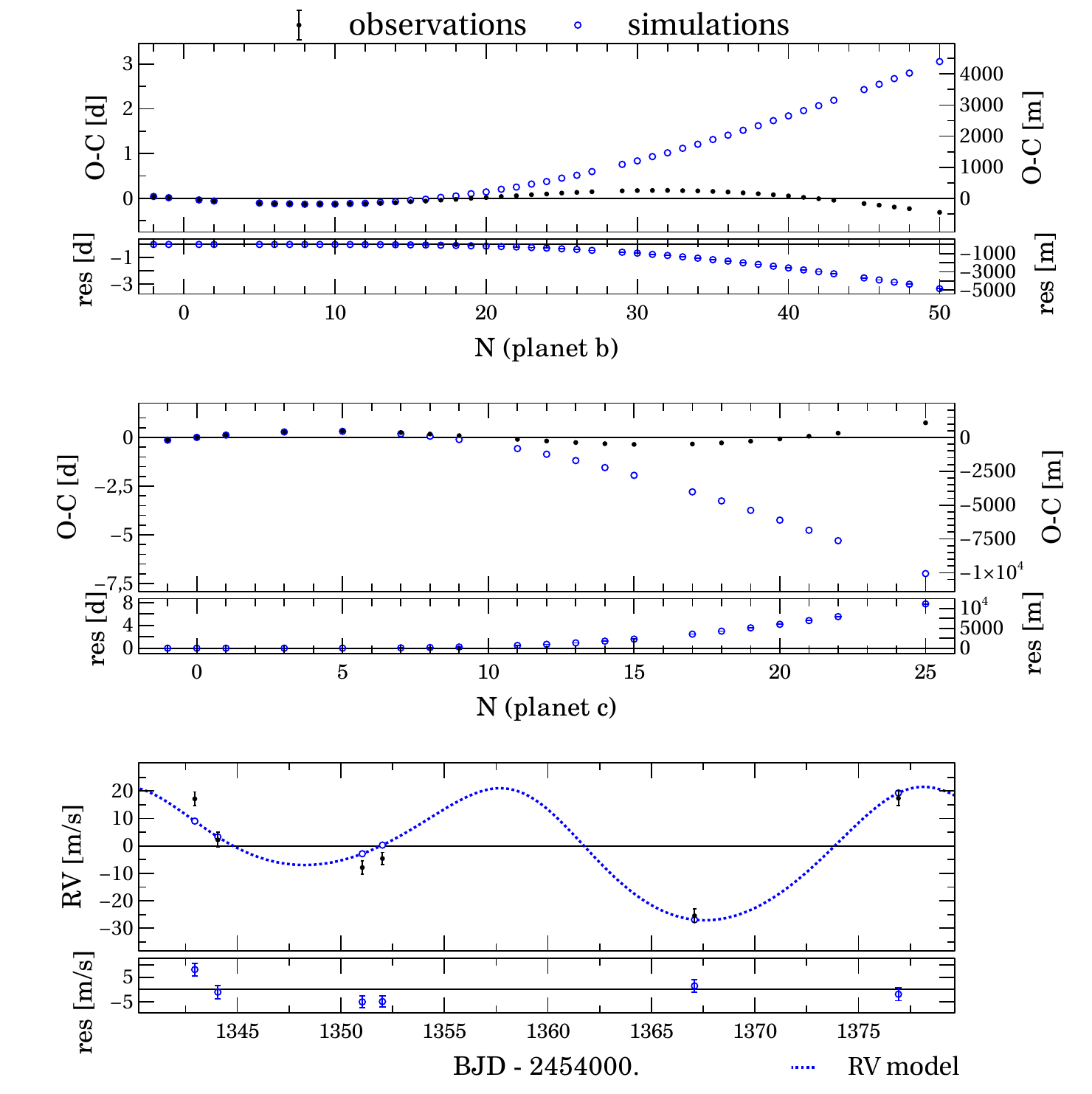}}
  \caption{
    Kepler-9 system: $O-C$ diagrams from the parameters obtained (solution K9-I in Table~\ref{TabPar}) with \texttt{TRADES} for the data from \citet{Holman2010Sci} extended to the twelve quarters of \textit{Kepler}.
    The simulations are compared with the $T_0$s from \citet{Mazeh2013arXiv}, while the RVs are from the discovery paper.
    The epoch of the transits ($N$ in x-axis) are calculated from the linear ephemeris from \citet{Mazeh2013arXiv}.
  }
  \label{FigInt01}
\end{figure}

As for the Kepler-11 system, we extended the analysis of Kepler-9 to the first twelve quarters of \textit{Kepler} data using the transit times from \citet{Mazeh2013arXiv}.
We did not find any transit time to discard when using the same criteria used for Kepler-11.
First of all, we extended the integration of the orbits of the planets from the solution K9-I to the twelve quarters (we did not fit any parameters in this simulation) and we compare the observed $T_0$s and RVs with the simulated ones.
In Fig.~\ref{FigInt01} it is clear that the simulation diverges quite soon from the observations.
We run a simulation with the \texttt{MERCURY} package with same initial parameters of \texttt{TRADES}, compared the resulting $O-C$ diagrams, and we found the same behavior.
Furthermore, we calculated the transit time differences between \texttt{TRADES} and \texttt{MERCURY}, and we found that the maximum absolute difference is about 12 seconds, which is really smaller than the error bars of the $T_0$s.\par

\begin{figure}[h]
  \resizebox{\hsize}{!}{\includegraphics{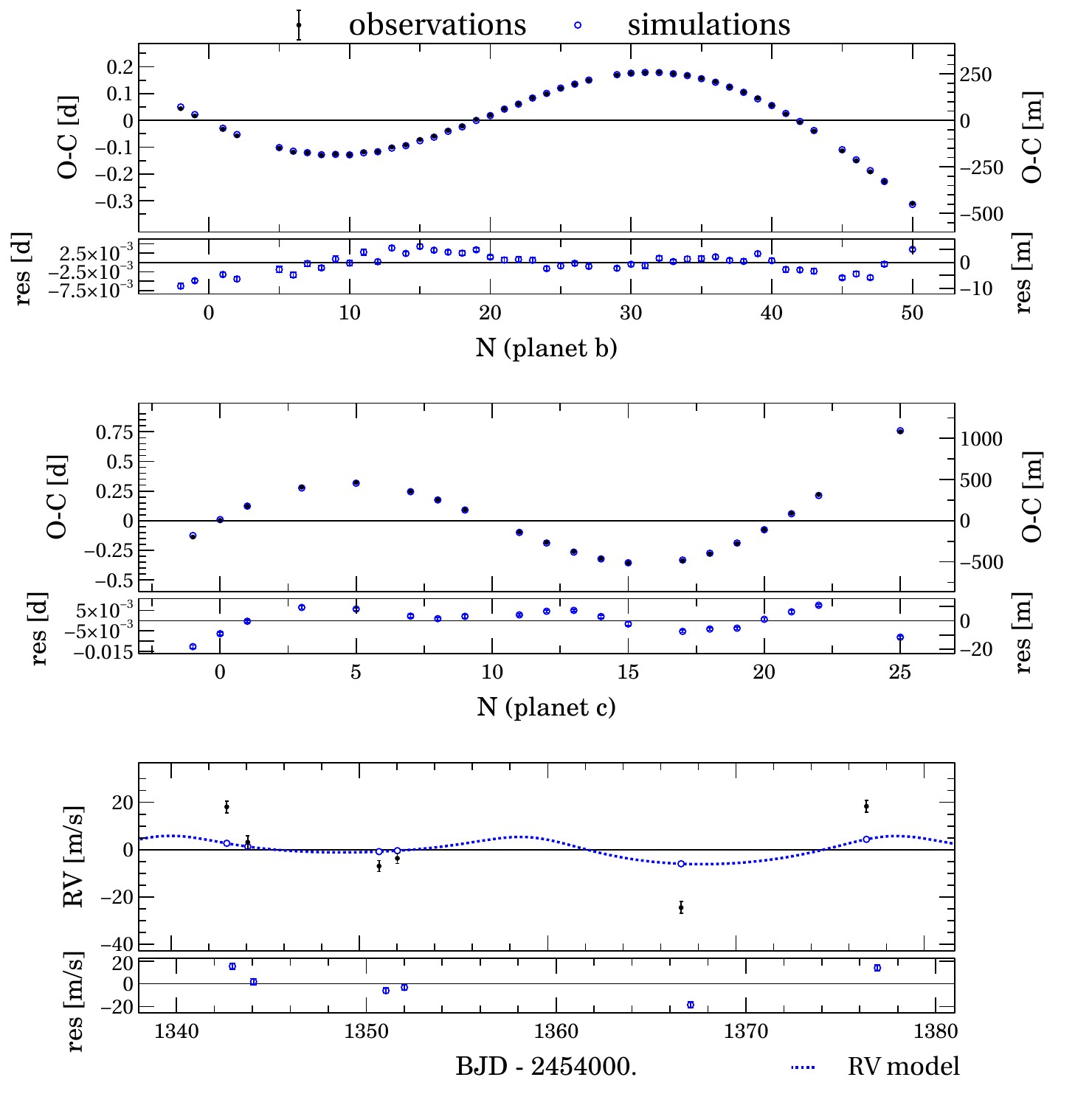}}
  \caption{
    Kepler-9 system: $O-C$ diagrams from the fit with \texttt{TRADES} for the data from the twelve quarters of \textit{Kepler}.
    $T_0$s from \citet{Mazeh2013arXiv}, while the RV are from the discovery paper.
    Initial parameters from the solution K9-I.
    $\chi^2_\mathrm{r} \approx 33.57$ for $62$ dof.
    Given the high value of the $\chi^2_\mathrm{r}$, we did not report the parameters of this solution.
  }
  \label{FigFit12}
\end{figure}

We considered the orbital solution K9-I in Table~\ref{TabPar}, and we ran a simulation on the $T_0$s of \citet{Mazeh2013arXiv} for the same first three quarters of \citet{Holman2010Sci}.
The six RV points are taken into account.
The fitted orbital solution has all the parameters agree with the solution K9-I.\\
Then, we fit all the 12 quarters with initial condition from solution K9-I.
The final $\chi^2_\mathrm{r}$ is $\sim 33$ (for $62$ dof).
The $O-C$ diagrams (Fig~\ref{FigFit12}) are fitted better than those in Fig.~\ref{FigInt01}, and the RV plot (bottom diagram in Fig~\ref{FigFit12}) shows a lower amplitude.
\par

To investigate the origin of this disagreement between observations and simulations when fitting 12 quarters (Figs.~\ref{FigInt01} and \ref{FigFit12}), we analyzed the $T_0$s by \citet{Mazeh2013arXiv} with N simulations, with each one fitting three adjacent quarters of data (we called it '3 moving quarters') and the six RV, in that we considered quarter 1 to 3, 2 to 4, and up to 10 to 12.
We set the parameters from solution K9-I as the initial parameters of each simulation.
We had good fits up to the simulation with quarters 6, 7, and 8;
the following simulations showed increased $\chi^2_\mathrm{r}$ ($>700$) that dropped to $\approx 44$ only for the last three moving simulation (quarters 10, 11, and 12).
In this analysis, we found that the bad fit starts when the solution K9-I diverges in Fig.~\ref{FigInt01}.
This could be an hint that the original solution determined by analyzing only the first three quarters of data is biased by the short time scale.

\subsection{Dynamical analysis without RV points}
\label{subNoRV}

Due to the high $\chi^2$ in Fig.~\ref{FigFit12}, we re-analyzed the Kepler-9 system in a different way.
We chose to run many simulations with \texttt{GA+LM} and \texttt{PSO+LM} on all 12 quarters with and without fitting the RV points.
We set quite wide bounds on the parameters; in particular, we set the masses to be bound between $10^{-6}\ M_\mathrm{Jup}$ and $1\ M_\mathrm{Jup}$ and the eccentricities between $0$ and $0.3$.
The best solution (K9-II) has been obtained with the \texttt{TRADES} mode \texttt{PSO+LM}) without an RV fit.
This solution has a $\chi^2_\mathrm{r}$ of about $1.44$ for $56$ dof (summary of the final parameters in Table~\ref{TabPar}).
The masses of solution K9-II are about $55\%$ of the masses published by \citet{Holman2010Sci}.
Furthermore, the eccentricities are smaller than the published ones \citep[calculated from the SOM of][]{Holman2010Sci} and of the order of $0.06$.
These small values of the masses and the eccentricities imply a RV semi-amplitude of about $16.11$ ms$^{-1}$, which is smaller than the one from the solution K9-I ($\sim 28.91$ ms$^{-1}$) that is extended to all 12 quarters.

\begin{figure}[h]
  \resizebox{\hsize}{!}{\includegraphics{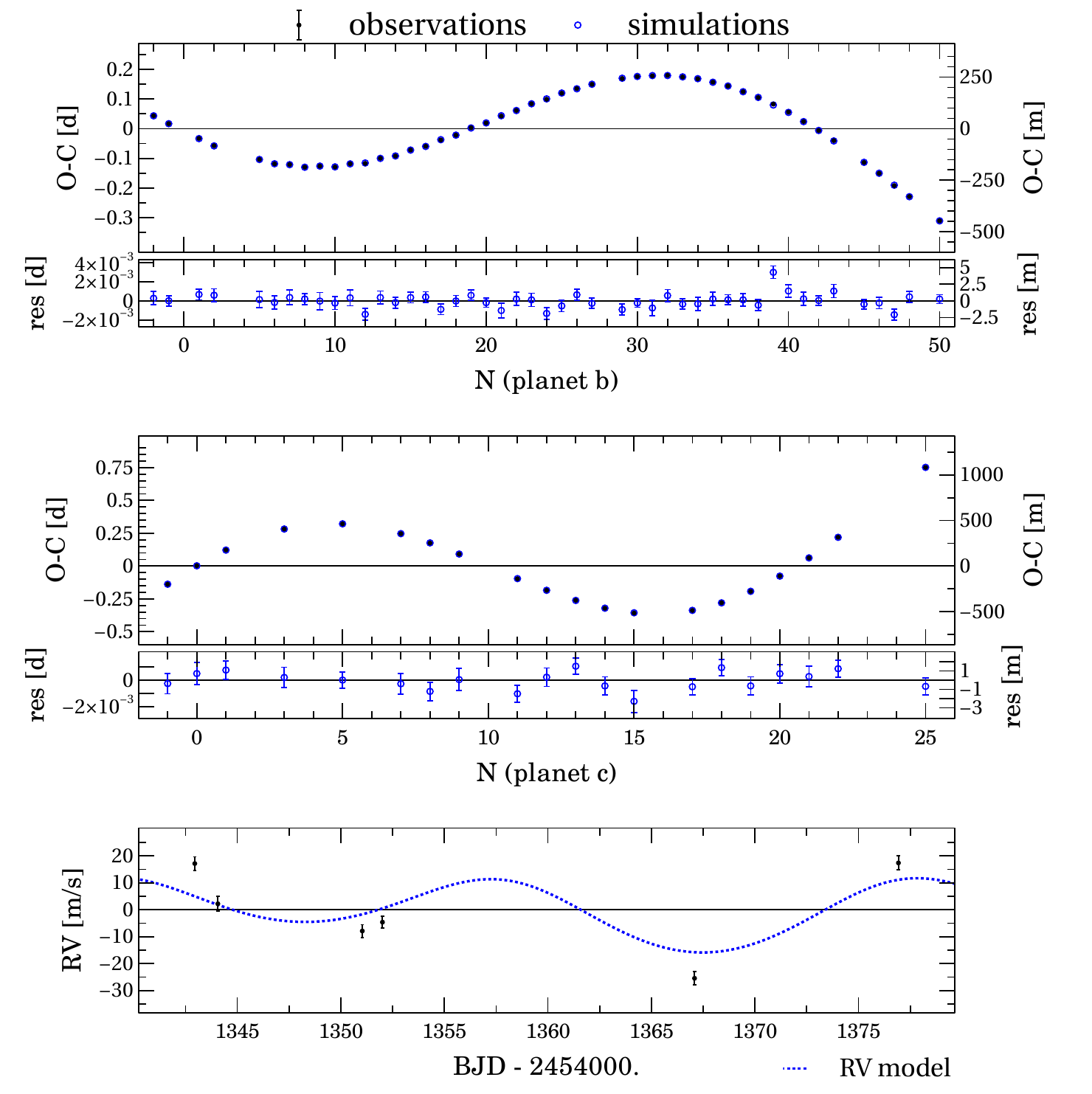}}
  \caption{
    $O-C$ diagrams (top and middle plot) for the solution K9-II in Table~\ref{TabPar} from the fit of the $T_0$s for the 12 quarters and neglecting the six RV points.
    Colors and markers as in Fig.~\ref{FigFit12}
    $\chi^2_\mathrm{r} \approx 1.44$ for $56$ dof.
    In the bottom panel, the RV model (blue dots) from the solution K9-II with over plotted RV observations (black dots with error-bars).
    We do not have the RV residuals from \texttt{TRADES}, because we have not fitted the RVs.
    The RV model has a lower RV semi-amplitude of about $12.80$ ms$^{-1}$ with respect to the RVs from the discovery paper.
  }
  \label{FigK9-II}
\end{figure}

\section{Summary}
\label{summary}

We have developed a program, \texttt{TRADES}, that simulates the dynamics of exoplanetary systems and that does a simultaneous fit of radial velocities and transit times data.\\
Analyzing a simulated planetary system, we have shown that \texttt{TRADES} can determine the parameters even from low precision data or for a rough guess of the initial orbital elements.
\par

We validated \texttt{TRADES} by reproducing the packed exoplanetary system Kepler-11.
The orbital parameters we determined agree with the values of the discovery paper and the recent analysis of 14 quarters by \citet{Lissauer2013ApJ}.
Furthermore, our analysis agrees with the results by \citet{Migas2012} that used a similar approach.
Our best simulation (K11-III) returned a value for the mass of the planet g of about $25\ M_{\oplus}$, which agrees within the error bars and the confidence interval proposed by \citet{Lissauer2013ApJ}.
Furthermore, all our simulations showed a final $\chi^2_\mathrm{r} \lesssim 2$, and the final mass of planet g agrees with the previous works.
\par

We reproduced the Kepler-9 system (without KOI-377.03), and we found that the parameters from the SOM of the discovery paper \citep{Holman2010Sci} cannot properly reproduce the $O-C$ diagram of Kepler-9c.
We tested the orbits and $O-C$ diagrams with an independent program, \texttt{MERCURY}, and we found the same result.
A difference of a few degrees in $\lambda$ for both planets is enough to explain the offset of about 300 minutes in the $O-C$ diagram.
We found the same results after analyzing the $T_0$s by \citet{Mazeh2013arXiv} that cover the same quarters of \citet{Holman2010Sci}.
\par

Extending the analysis, we found that the original solution is not compatible to the whole set of data from the 12 quarters by \citet{Mazeh2013arXiv}.
It shows a divergence of the simulated $O-C$ compared to the observed one (Fig.~\ref{FigInt01}).
The solution K9-I, as obtained with the same temporal baseline of the discovery paper cannot explain the observed transit time for the whole set of $T_0$s.\\
Only using the combination of a `quasi' global optimized search algorithm, such as genetic \texttt{PIKAIA} or particle swarm (\texttt{PSO}), with the \texttt{LM} algorithm, it has been possible to improve the fit on all 12 quarters but only if we neglect the six RV points.
With this approach, we have found a new solution (K9-II) with a $\chi^2_\mathrm{r} \approx 1.4$ for $56$ dof.
This solution lead to mass values that are about $55\%$ of the mass values given in the discovery paper and smaller eccentricities.
Due to these values, our RV model has a smaller semi-amplitude of about $12.80$ ms$^{-1}$ if compared to the observations.\\
We need to better study this system by, for example, obtaining more RV points because it can shed light on the issue of the different masses of the exoplanets calculated from RV data and from the TTV \citep[for a similar case see][]{Masuda2013}.
Follow-up transit observations with CHEOPS will extend the time coverage and are advisable.\par

In both \textit{Kepler} systems we carried out the analysis of twelve quarters using the transit times calculated by \citet{Mazeh2013arXiv} using an automated algorithm.
We point out that it would be advisable to analyze the light curves of the KOIs that show TTV signals and recompute the transit times of the planets in more detail.
\par

It is known that the \texttt{LM} algorithm cannot return reliable errors (with physical meaning) in presence of correlated errors and complex parameter spaces.
For some parameters, the bootstrap analysis returned small intervals of confidence.
A possible explanation is that the parameter distributions in the bootstrap analysis are limited to values close to those found by the \texttt{LM} algorithm.
This is probably due to a strong selection effect of the forest of minima in the $\chi^2$ space.
Furthermore, for the Kepler-9 case, the measurement errors have tiny effect compared to the TTV signal that dominates the distribution of the parameters.
\par

In the near future, we add a Monte-Carlo-Markov-Chain (MCMC) algorithm \citep[or an another Bayesian algorithm, e.g.,\ multiNEST,][]{multiNEST2009} in \texttt{TRADES} to perform parameter estimation and model selection using a Bayesian approach.
The idea is to provide the initial parameters to the Bayesian algorithm via the \texttt{LM} algorithm or use a grid search.
Furthermore, we plan to include the transit duration in the fitting procedure to put more constraint on the parameter determination.
\par

Recently, the European Space Agency (ESA) adopted the space S-class mission CHEOPS \citep{CHEOPS2013}.
The launch of this satellite is foreseen by December 2017.
CHEOPS will characterize the structure of Neptune- and Earth-like exoplanets monitoring their transits.
The CHEOPS targets will be stars hosting planets that are known via accurate Doppler and ground-based surveys.
We plan to optimize \texttt{TRADES} for CHEOPS mission to dynamically characterize observed exoplanetary systems and to analyze possible TTV detections.
\par

The work by \citet{Mazeh2013arXiv} provided a list of exoplanetary systems that show TTV signals.
We are currently selecting a wide sample of these systems that would be suitable for an analysis with the program \texttt{TRADES}, such as Kepler-19.
We plan to make a self-consistent analysis of the raw \textit{Kepler} data, determine the transit times, and combine them with available RV data.
\par

\texttt{TRADES} will be publicly release as soon as possible.
Those interested can send an email to the first author of this work, and a copy of the code will be provided.
At the moment the documentation is still in a preliminary form, but we are willing to provide all information needed.
\par

\begin{acknowledgements}
  We thank the referee David Nesvorn\'y for his careful reading and useful comments and suggestions.
  We acknowledge support from Italian Space Agency (ASI) regulated by Accordo ASI-INAF n. 2013-016-R.0 del 9 luglio 2013.
\end{acknowledgements}

\end{document}